\documentclass[aps,prd,preprint,superscriptaddress,nofootinbib]{revtex4-1}

\usepackage{graphicx}
\usepackage{booktabs}
\usepackage{hyperref}
\usepackage{amssymb,amsmath,amsfonts}
\usepackage{color}
\usepackage[utf8]{inputenc}
\usepackage{subfig}
\usepackage {diagbox}

\newcommand{\beq}{\begin{eqnarray}}
\newcommand{\eeq}{\end{eqnarray}}
\newcommand{\p}{\partial}

\newcommand{\vs}[1]{\vspace{#1 mm}}
\newcommand{\hs}[1]{\hspace{#1 mm}}
\newcommand{\bpm}{\begin{pmatrix}}
\newcommand{\epm}{\end{pmatrix}}
\newcommand{\Z}{\mathbb{Z}}
\newcommand{\R}{\mathbb{R}}
\newcommand{\C}{\mathbb{C}}

\newcommand{\D}{\mathcal D}

\newcommand{\ba}{\left(\begin{array}}
\newcommand{\ea}{\end{array} \right)}

\begin{document}

\title{Resurgence and semiclassical expansion \\in two-dimensional large-$N$ sigma 
models}

\author{Hiromichi Nishimura}
\affiliation{Department of Physics \& Research and Education Center for Natural Sciences, Keio University, Hiyoshi 4-1-1, Yokohama, Kanagawa 223-8521, Japan}
\affiliation{Research Center for Nuclear Physics (RCNP), Osaka University, Osaka 567-0047, Japan}

\author{Toshiaki Fujimori}
\affiliation{Department of Physics \& Research and Education Center for Natural Sciences, Keio University, Hiyoshi 4-1-1, Yokohama, Kanagawa 223-8521, Japan}

\author{\\ Tatsuhiro Misumi}
\affiliation{Department of Physics, Kindai University, Osaka 577-8502, Japan \\ \vspace{2cm}}
\affiliation{Department of Physics \& Research and Education Center for Natural Sciences, Keio University, Hiyoshi 4-1-1, Yokohama, Kanagawa 223-8521, Japan}

\author{Muneto Nitta}
\affiliation{Department of Physics \& Research and Education Center for Natural Sciences, Keio University, Hiyoshi 4-1-1, Yokohama, Kanagawa 223-8521, Japan}

\author{Norisuke Sakai}
\affiliation{Department of Physics \& Research and Education Center for Natural Sciences, Keio University, Hiyoshi 4-1-1, Yokohama, Kanagawa 223-8521, Japan}

\begin{abstract}
The resurgence structure of the 2d $O(N)$ sigma model 
at large $N$ is studied with a focus on 
an IR momentum cutoff scale $a$ that regularizes
IR singularities in the semiclassical expansion. 
Transseries expressions for condensates and correlators are derived as series of the dynamical scale 
$\Lambda$ (nonperturbative exponential) 
and coupling $\lambda_{\mu}$ renormalized 
at the momentum scale $\mu$. 
While there is no ambiguity when $a > \Lambda$, 
we find for $a < \Lambda$ that
the nonperturbative sectors have new imaginary ambiguities
besides the well-known renormalon ambiguity 
in the perturbative sector. 
These ambiguities arise 
as a result of an analytic continuation of 
transseries coefficients to small values of the IR cutoff $a$ 
below the dynamical scale $\Lambda$. 
We find that the imaginary ambiguities are 
cancelled each other 
when we take all of them into account. 
By comparing the semiclassical expansion 
with the transseries for the exact large-$N$ result, 
we find that some ambiguities vanish 
in the $a \rightarrow 0$ limit 
and hence the resurgence structure changes
when going from the semiclassical expansion 
to the exact result with no IR cutoff. 
An application of our approach 
to the ${\mathbb C}P^{N-1}$ sigma model is also discussed. 
We find in the compactified model with the $\Z_N$ twisted boundary condition that the resurgence structure changes discontinuously as the compactification radius is varied. 
\end{abstract}

\maketitle
\newpage
\tableofcontents

\section{Introduction}
\label{sec:Introduction}

Quantum field theory (QFT) is arguably 
the pinnacle of human knowledge of the microscopic world.  
It is the language used 
in the standard model of particle physics 
and condensed matter physics. 
Despite its success in many areas of physics, 
there is no rigorous definition of QFT in continuum yet. 
There has recently been, however, a considerable effort 
to use the resurgence theory \cite{Ec1,Pham1,BH1,Howls1,DH1,Costin1,Sauzin1,Sauzin2,Marino:2012zq,Dorigoni:2014hea,Aniceto:2018bis} 
to study QFT and to give a continuum definition of it. 
There have been particularly a lot of progress 
in quantum mechanics:
the resurgent structure in double-well and periodic potentials \cite{Bender:1969si, Bender:1973rz, Brezin:1977ab, Lipatov:1977cd, Bogomolny:1980ur, ZinnJustin:1981dx, ZinnJustin:1982td, ZinnJustin:1983nr},
the valley method \cite{Aoyama:1991ca,Aoyama:1994sk,Aoyama:1995ca, Aoyama:1997qk,Aoyama:1998nt},
exact quantization conditions and constructive resurgence \cite{ZinnJustin:2004ib, ZinnJustin:2004cg, Jentschura:2010zza, Jentschura:2011zza, Jentschura:2004jg, Dunne:2013ada, Basar:2013eka,Dunne:2014bca,Misumi:2015dua,Gahramanov:2015yxk,Dunne:2016qix},
bion cancellation mechanism 
and the other cancellation mechanics \cite{Behtash:2015zha,Behtash:2015loa,Fujimori:2016ljw, Sulejmanpasic:2016fwr,Dunne:2016jsr,Kozcaz:2016wvy,Serone:2016qog,Basar:2017hpr,Fujimori:2017oab,Serone:2017nmd,Behtash:2017rqj,Costin:2017ziv,Alvarez:2017sza,Fujimori:2017osz,Sueishi,Ito:2018eon,  Behtash:2018voa, Pazarbasi:2019web, Sueishi:2020rug}. 
The resurgent structure of 
two-dimensional quantum field theories 
has been also investigated:
sigma models \cite{Dunne:2012ae, Dunne:2012zk,Misumi:2014jua,Misumi:2014bsa,
Misumi:2014rsa,Nitta:2014vpa,Nitta:2015tua,Behtash:2015kna,Dunne:2015ywa,Misumi:2016fno,Sulejmanpasic:2016llc,Fujimori:2018kqp,Ishikawa:2019tnw,Yamazaki:2019arj,Ishikawa:2020eht,Morikawa:2020agf},
principle chiral models \cite{Cherman:2013yfa,Cherman:2014ofa,Demulder:2016mja},
and 2D Yang-Mills theory \cite{Okuyama:2018clk}. 
This paper is a continuation of such an effort 
to study QFT in the framework of resurgence theory. 

In most of continuum quantum field theories, 
the exact solution is not known, 
and one of the most reliable approaches 
may be the semiclassical method,
which uses a transseries 
with powers of nonperturbative exponential 
$e^{-A/\lambda_\mu}~(A: const.)$ 
and divergent power series in coupling $\lambda_\mu$, 
where the subscript $\mu$ indicates that 
the coupling constant is renormalized at the momentum scale $\mu$. 
Schematically, the expectation value of an observable $\mathcal{O}$ may be written as  
\begin{equation}
\left< \mathcal{O} \right>
=
\sum^{\infty}_{l=0} e^{-l A/ \lambda_\mu} C_l(\lambda_\mu), \hs{10}
 C_l(\lambda_\mu) = 
\sum^{\infty}_{n=0} c_{\left(l,n \right)} \lambda_\mu^{n} . 
\label{sc_expansion}
\end{equation}
The $l=0$ sector is the usual perturbative expansion 
around the trivial vacuum, 
while the $l>0$ sector is a small-coupling expansion 
around the $l$-th nonperturbtive background configuration. 
The latter contains an essential singularity $e^{-1/\lambda_\mu}$ 
in the complex $\lambda_\mu$ plane, 
and therefore it is nonperturbative in nature. 
Generically $C_l(\lambda_\mu)$ are divergent asymptotic series: 
the sum over all $n$ is neither convergent nor Borel summable. 
Typically the expansion coefficients are 
factorially divergent  $c_{(l,n)} \sim n!$, 
and this gives rise to singularities in the Borel plane.
These Borel singularities are sometimes located 
on the real positive axis in the Borel plane, 
which leads to an imaginary ambiguity of 
the resultant Borel resummation. 
They are expected to cancel with each other when all the ambiguities (including those from discontinuous jumps of Stokes constants) are taken into account. 
As a consequence, it connects the perturbative and nonperturbative contributions in the physical quantities via the imaginary ambiguities. 
This is one of the final goals of the application of the resurgence theory: Resurgence theory attempts to find out a nontrivial relation between the perturbative and nonperturbative sectors that enables us to extract the nonperturbative information just from the perturbative series through the Borel resummation and its imaginary ambiguity.

The Borel singularities of 
the factorially divergent perturbative series 
can be classified into two classes. 
The first one comes from 
the proliferation of the number of Feynman diagrams. 
Ambiguities associated with this type of Borel singularities 
are found to be cancelled 
by nonperturbative contributions of 
saddle point configurations such as a bion 
(a fractional instanton-anti-instanton composite) 
\cite{Dunne:2012ae, Dunne:2012zk}. 
The second one is the ``renormalon" type \cite{tHooft:1977xjm,Beneke:1998ui}, 
whose imaginary ambiguities 
can not naively be cancelled 
by those of saddle point configurations. 
In general, the renormalon type ambiguity is 
related to contributions of a selected set of the Feynman diagrams with loop momentum integrations 
written in terms of the renormalized coupling constant. 
They also contribute a factorially divergent series, 
whose Borel resummation gives an imaginary ambiguity. 
In particular, ambiguities of renormalon type remain
even at large $N$, 
while those associated with the proliferation of diagrams 
vanish in the large-$N$ limit. 
There is a conjecture that the renormalon 
can be identified as a certain semiclassical object (e.g. bion) 
in the compactified spacetime 
in the Euclidean path integral formulation 
\cite{Dunne:2012ae, Dunne:2012zk}, 
on which both the affirmative 
and negative arguments have been developed 
\cite{Fujimori:2018kqp,Ishikawa:2019tnw,Yamazaki:2019arj,
Ishikawa:2020eht,Morikawa:2020agf}.
The way how the renormalon ambiguity is cancelled out 
by nonperturbative contributions is still an open question.  

In this paper, we would like to shed light 
on these issues by revisiting the $O(N)$ non-linear sigma model 
in two dimensions at large $N$. 
The resurgence structure and renormalons in the $O(N)$ sigma model
have been studied by using method such as 
the large-$N$ analysis and integrability 
\cite{David:1982qv,David:1983gz,David:1985xj,Novikov:1984ac,Beneke:1998eq,Volin:2009wr,Marino:2019eym,Abbott:2020mba,DiPietro:2021yxb,Schubring:2021hrw,Marino:2021dzn,Bajnok:2021dri,Bajnok:2021zjm}.
The model is an asymptotically free theory with a mass gap,
and many important properties 
can be exactly solved at large $N$.  
In this paper, we mainly focus on the condensate
$\left< \delta D^2 \right>$ 
of fluctuations of an auxiliary field $D(x)$, 
as the perturbative expansion is known to have a renormalon.  
Intended to simulate the massless perturbation theory, 
we compute the transseries coefficients $c_{(l,n)}$ 
using the transseries expansion of the momentum integrand,
which can be exactly determined in the large-$N$ limit,  
in powers of the nonperturbative exponential 
$e^{-2\pi/\lambda_p} = \Lambda/p$ and 
the coupling constant $\lambda_p = 2\pi/\log(p/\Lambda)$
renormalized at momentum scale $p$. 
In this approach, we encounter IR divergences 
like in the massless perturbation theory, 
since the model is perturbatively massless
even though there is a nonperturbative mass gap. 
We obtain transseries and examine both renormalons 
and IR divergences in the model with our semiclassical ansatz. 

In the rest of this section, 
we outline the paper and highlight our main results. 
In Sec.\,\ref{sec:renormalon}, 
we briefly review how renormalon type ambiguities appear 
in a generic asymptotically free theory. 
In Sec.\,\ref{sec:Review}, 
we review the $O(N)$ sigma model at large $N$ 
and write down the exact large-$N$ expression 
of the condensate $\langle \delta D^2 \rangle$. 
The asymptotic series of the exact result 
contains a divergent perturbative expansion, 
which is not Borel summable. 
The singularity in the Borel plane gives an renormalon type ambiguity of order $\Lambda^4$. 
Since the exact expression is real and unambiguous, 
such a renormalon ambiguity must be cancelled. 
The question we would like to address is 
how the cancellation of the renormalon occurs 
in the context of the semiclassical expansion.  
In Sec.\,\ref{sec:SemiclassicalForm}, 
we argue that the expected form 
for the semiclassical (s.c.) expansion 
of the condensate can be written as
\begin{equation}
\left< \delta D^2 \right> \
\underset{\rm s.c.}{\sim} \
\sum^{\infty}_{l=0} e^{-4 \pi l / \lambda_\mu}  \sum^{\infty}_{n=0} 
\lambda^{n+1}_\mu \int \frac{d^d p}{(2\pi)^d}  
h_{\left(  l, n \right) } (p),
\label{CF_sc_expansion}
\end{equation}
in $d$ dimensions. 
It is beyond the scope of this paper 
to derive the semiclassical expansion  
by explicitly computating the perturbation series around 
the vacuum and nontrivial backgrounds. 
Instead, we simply deduce 
the semiclassical form (\ref{CF_sc_expansion}) 
from the exact solution at large $N$. 
In our semiclassical ansatz, 
the momentum integral becomes 
IR divergent for higher orders in $l$
since the model is perturbatively massless. 
We therefore need to introduce 
an IR momentum cutoff $a$ ($|p|>a$).
In each sector labeled by $l$ (of order 
$e^{-4 \pi l / \lambda_\mu}=(\Lambda/\mu)^{2l}$), 
the sum over $n$ can contain a factorially divergent series, 
but their Borel resummations have no 
imaginary ambiguities at any $l$
when the IR cutoff is larger 
than the dynamical scale $a > \Lambda$. 
In Sec.\,\ref{sec:Cancellation}, 
we show that when the IR cutoff is small, $0<a < \Lambda $, 
the imaginary ambiguities arise 
at order $\Lambda^0$, $\Lambda^4$, and $\Lambda^8$. 
This shows that the presence of imaginary ambiguities 
depends on the IR cutoff $a$.  
In Sec.\,\ref{sec:PerturbativeExpansion}, 
we investigate the origin of the imaginary ambiguities 
by explicitly computing the semiclassical expansion 
up to order $\Lambda^8$.
We first identify that the ambiguity at $\Lambda^0$ 
is the well-known renormalon in perturbation theory.   
We then show that this imaginary ambiguity is cancelled 
by the combined imaginary ambiguities 
that come from order $\Lambda^4$ and $\Lambda^8$ 
in the semiclassical ansatz, 
and not only from order $\Lambda^4$ 
as previously known for the transseries of 
the exact large-$N$ result with $a=0$.  
This is one of our main results in this paper. 
In Sec.\,\ref{sec:Comparison}, 
we examine the result of the semiclassical ansatz 
by comparing it with the exact result with $a=0$. 
We find that the ambiguities at order $\Lambda^8$ 
can be understood as the result of an analytic continuation 
in $\lambda_a$ below the dynamical mass $a<\Lambda$, 
where $\lambda_a<0$. 
We also find how to obtain $a\to 0$ limit.
In Sec.\,\ref{sec:correlation_function}, 
we generalize the discussion to the correlation function
$\langle \delta D(x) \delta D(0) \rangle$. 
In Sec.\,\ref{sec:OtherModels}, 
we discuss the generalization to 
the ${\mathbb C}P^{N-1}$ sigma model
including the case of the $\Z_N$ 
twisted periodic boundary condition. 
We show that the resurgence structure 
changes discontinuously 
when each Kaluza-Klein mass (Matsubara frequency) 
$2\pi n/L~(n \in \Z)$ 
becomes smaller than $\Lambda$ 
as we vary the compactification period $L$.  
Sec.\,\ref{sec:Conclusions} gives conclusion and discussion.

\section{Renormalons in asymptotically free theories}
\label{sec:renormalon}
In this section, let us briefly review how renormalons 
appear in two-point functions and condensates 
in asymptotic free theories. 
Suppose that there is a renormalized coupling constant 
$\lambda_\mu$
depending on the renormalization scale $\mu$ 
as determined by the renormalization group equation
\beq
\mu \frac{\p}{\p \mu} \frac{2\pi}{\lambda_\mu} = \beta(\lambda_\mu),
\eeq
where $\beta(\lambda_\mu)$ is the beta function, 
whose expansion takes the form 
\beq
\beta(\lambda_\mu) = \beta_0 + \beta_1 \lambda_\mu + \beta_2 \lambda_\mu^2 + \cdots. 
\label{eq:beta_function}
\eeq
We assume that 
the first coefficient is positive $\beta_0 > 0$
so that the coupling constant $\lambda_\mu$ 
becomes small for $\mu \rightarrow \infty$, 
i.e. the model is asymptotically free\footnote{
In this paper, the symbol $\lambda_\mu$ stands for 
the renormalized 't Hooft coupling, 
for which $\beta_0 = 1$ 
in the $O(N)$ and $\C P^{N-1}$ sigma models.}.
We define the dynamically generated scale $\Lambda$ 
as the scale at which the renormalized coupling constant diverges 
\beq
\lambda_\mu (\mu/\Lambda) ~\overset{\mu \rightarrow + \Lambda}{\longrightarrow} \infty.
\eeq
Note that $\lambda_\mu$ is a function of $\mu/\Lambda$.
In the following, we assume that 
the model has no parameter with a mass scale 
other than $\Lambda$. 

Suppose that we are interested in the two-point function 
of a local operator $\mathcal O$
\beq
\langle \mathcal O(x) \mathcal O(0) \rangle = 
\int \frac{d^d p}{(2\pi)^d} e^{ip \cdot x} \Delta(p), 
\label{eq:2pt_function_general}
\eeq
where $\Delta(p)$ is 
the (Euclidean) momentum space propagator, 
that is, the correlation function of  
the Fourier transform $\tilde{\mathcal O}(p)$ of $\mathcal O(x)$
\beq
\langle \tilde{\mathcal O}(p) \tilde{\mathcal O}(p') \rangle = \delta(p+p') \Delta(p).
\eeq
We assume that $\Delta(p)$ has 
no singularity as a function of $p_i \in \R^d$
and the integral \eqref{eq:2pt_function_general} is well-defined. 
In particular, the regularity of the propagator 
at the origin implies that there is a mass gap in this model.  
The semiclassical expansion with respect to
the renormalized coupling constant $\lambda_\mu$ would give
the following transseries expression for the propagator
\beq
\Delta(p) = p^{[\Delta]} \, \sum_{l=0}^\infty \exp \left( -\frac{2\pi\sigma_l}{\lambda_\mu} \right) h_l \left(\frac{p}{\mu},\lambda_\mu \right), \hs{5}
h_l \left( \frac{p}{\mu},\lambda_\mu \right) = \sum_{n=0}^\infty a_{ln} \! \left( \frac{p}{\mu} \right) \, \lambda_\mu^n,
\label{eq:2pt_momentum_transseries_0}
\eeq
where $[\Delta]$ is the mass dimension 
of the propagator $\Delta(p)$ related to that of the operator $\mathcal O$ as $[\Delta] = 2[\mathcal O]-d$, 
$\sigma_l~(0 < \sigma_1 < \sigma_2 < \cdots)$ 
are the nonperturbative exponents,
$a_{ln}(p/\mu)$ are the $n$-th expansion coefficients 
in the $l$-th nonperturbative sector.
In the following, we call expressions like \eqref{eq:2pt_momentum_transseries_0} ``semiclassical ansatz" 
since transseries of this type would be the expected form 
obtained in the semiclassical expansion: 
the $l$-th sector is the contribution 
from the $l$-th saddle point configuration
characterized by the action $S_l \propto \sigma_l$
and $h_l$ stands for the power series obtained through the perturbative expansion around the saddle point configuration. 
For simplicity, we assume that the semi-classical expansion 
for the propagator \eqref{eq:2pt_momentum_transseries_0} 
is convergent for $p > \mu > \Lambda$.
This assumption is true in the large-$N$ case discussed below. 
Using the renormalization group, 
we can change the renormalization scale from $\mu$ to $p$. 
Then the transseries for the propagator becomes
\beq
\Delta(p) = 
p^{[\Delta]} \, \sum_{l=0}^\infty \exp \left(-\frac{2\pi\sigma_l}{\lambda_p} \right) \tilde h_l(\lambda_p) , \hs{10}
\tilde h_l(\lambda_p) = \sum_{n=0}^\infty b_{ln} \, \lambda_p^n.
\label{eq:2pt_momentum_transseries_1}
\eeq
In this expression, 
the expansion coefficients $b_{ln}$ 
are constants without $p$-dependence 
since there is no other mass scale. 
Note that each term in this transseries is 
singular at $p=\Lambda$ 
due to the singularity of the renormalized coupling $\lambda_p$.
The singularity structure becomes more manifest 
by expanding the renormalized coupling constant $\lambda_p$
in powers of the one-loop coupling constant $\lambda_p'$ as\footnote{
In the large-$N$ sigma model discussed below, 
$\lambda_p'$ is identified with $\lambda_p$ 
since the correction is subleading in the large-$N$ limit.
}
\beq
\lambda_p \left(p \right) = \lambda_p' - \frac{\beta_1}{\beta_0}  {\lambda_p'}^2 \log \frac{4\pi}{\lambda_p'} + \cdots 
~~~\mbox{with}~~~ \lambda_p' = \frac{2\pi}{\beta_0 \log p/\Lambda} .
\eeq
Then, the transseries \eqref{eq:2pt_momentum_transseries_1} would be rewritten as
\beq
\Delta(p) = p^{[\Delta]} \, \sum_{l=0}^\infty \left( \frac{\Lambda}{p} \right)^{\beta_0 \sigma_l} f_l (\lambda_p'), \hs{7}
f_l (\lambda_p')= {\lambda_p'}^{\alpha}\sum_{n=0}^\infty c_{ln}{\lambda_p'}^n, 
\label{eq:transseries_Lambda}
\eeq
where $\alpha = 2\pi \sigma_l \beta_1/\beta_0$ and 
$c_{ln}$ are functions of 
$\log 4\pi/\lambda_p'$.  
Because of the asymptotic freedom, 
this transseries expression can also be viewed as 
the large-$p$ expansion of the propagator. 
If the function $f_l(\lambda_p')$ is divergent 
in the limit $\lambda_p' \rightarrow \infty$, 
the $l$-th term of the transseries 
\eqref{eq:transseries_Lambda} has a singularity at $p=\Lambda$
originating from that of the renormalized coupling $\lambda_p$.  
Due to this singularity at $p = \Lambda$, 
each term in the transseries for the two point function  
\beq
\langle \mathcal O(x) \mathcal O(0) \rangle_a = \sum_{l=0}^\infty \int_{|p|>a} \frac{d^dp}{(2\pi)^d} \, e^{ip \cdot x} \, p^{[\Delta]} \left( \frac{\Lambda}{p} \right)^{\beta_0 \sigma_l} f_l (\lambda_p'), 
\eeq
has an ambiguity depending on the regularization
if the singularity at $p=\Lambda$ is contained 
in the integration domain $|p|>a$. 
Here we have introduced an IR cutoff scale $a$ 
to regularize the singularity at $p=0$ 
in the integration for each term in the transseries 
\eqref{eq:transseries_Lambda}.
Such an IR cutoff is always necessary 
in the semi-classical computation
in a perturbatively massless model 
even though there is a dynamically generated mass gap\footnote{Instead of the IR momentum cutoff, 
we may introduce other deformations 
such as mass deformations, chemical potentials 
or background fields like the $\Omega$-background.}. 
On the other hand, the existence of the mass gap guarantee 
that there is a well-defined limit $a \rightarrow 0$ 
for the full two point function. 
Since the propagator itself has no singularity, 
all the ambiguities cancel out 
and the $a \rightarrow 0$ limit is regular. 

We can associated the ambiguity from the singularity at $p=\Lambda$ 
with a singularity in the Borel plane. 
For simplicity, let us focus on the case $x \rightarrow 0$, 
where the two point function reduces to the condensate
\beq
\langle \mathcal O(x) \mathcal O(0) \rangle_a ~\rightarrow~ \langle \mathcal O(0)^2 \rangle_{\tilde a,a} = \int_{a<|p|<\tilde a} \frac{d^dp}{(2\pi)^d} \, \Delta(p),
\label{eq:general_condensate}
\eeq
where we have introduced another cutoff scale $\tilde a$ to regularize the UV divergence. 
Assume that the transseries for the propagator \eqref{eq:transseries_Lambda}
has the following Borel resummed form
\beq
\Delta(p) = 2 \pi p^2 \sum_{l=0}^\infty \left( \frac{\Lambda}{p} \right)^{\beta_0 \sigma_l} \int_0^\infty dt \, \left( \frac{\Lambda}{p} \right)^{t} P_l(t). 
\eeq
By a change of variable and some manipulation (see Appendix \ref{appendix:general_Borel}), 
we can rewrite the condensate as
\beq
 \langle \mathcal O(0)^2 \rangle_{\tilde a,a} = C \, \mu^{2[\mathcal O]} \sum_{l=0}^\infty \left( \frac{\Lambda}{\mu} \right)^{\beta_0 \sigma_l} \int_0^\infty dt \, \left( \frac{\Lambda}{\mu} \right)^{t} B_l(t)
~~ \mbox{with} ~~ C = \frac{d \log \frac{\mu}{\Lambda}}{(4\pi)^\frac{d}{2} \Gamma(d/2+1)}.
\label{eq:Borel_expansion}
\eeq
The function $B_l(t)$ are given by\footnote{This ``Borel transform" $B_l(t)$ has a $\lambda_\mu'$-dependence. The standard coupling independent Borel transform will be denoted as $\mathcal B_l(t)$ in Sec.\,\ref{sec:PerturbativeExpansion}. 
We will use the same symbol $t$ for the variables of 
$B_l(t)$ and $\mathcal B_l(t)$ 
although they are not exactly identical.
}
\beq
B_l(t) = \frac{1}{e_l} \left[ \left( \frac{\mu}{a} \right)^{e_l} f_l \left( \frac{e_l \lambda_\mu'}{t+t_a} \right) - \left( \frac{\mu}{\tilde a} \right)^{e_l} f_l \left( \frac{e_l \lambda_\mu'}{t+t_{\tilde a}} \right) \right],
\eeq
where
\beq
e_l = \beta_0 \sigma_l-2[\mathcal O], \hs{10} 
t_p = e_l \, \frac{\log p/\Lambda}{\log \mu/\Lambda}.
\eeq 
If $f_l(\lambda_p')$ is divergent 
in the limit $\lambda_p' \rightarrow \infty~(p \rightarrow \Lambda)$, 
the Borel transform $B_l$ has a singularity at $t = - t_a$. 
This singularity is on the integration contour and 
gives rise to an ambiguity if $t_a$ is negative, 
i.e. the IR cutoff scale $a$ is smaller than 
the dynamically generated scale $\Lambda$. 
We can see that this singularity 
and the corresponding ambiguity do not vanish 
even in the large-$N$ limit. 
For example, 
the singularity and the corresponding ambiguity is
independent of $N$ in the $O(N)$ sigma model 
since $\beta_0=1,\,\sigma_l=2l$.
The factorial divergence of the perturbation series 
can also be seen from the fact that 
the Taylor expansion of $B_l(t')$ around $t=0$ 
has a finite radius of convergence due to the singularity.
In this way, the singularity of the renormalized coupling constant at $p=\Lambda$ results in renormalon type ambiguities.
In the next section, 
we will explicitly examine these renormalon ambiguities
in the $O(N)$ sigma model in the large-$N$ limit. 

\section{$O(N)$ sigma model at large $N$}
\label{sec:Review}

In this section, 
we give a brief review of the $O(N)$ sigma model 
in two dimensions at large $N$ 
in order to establish our notations 
and to write down the exact expression 
for the correlation functions and the condensate.
More comprehensive reviews on this subject can be found in 
Refs.~\cite{Novikov:1984ac,Beneke:1998ui,Marino:2015yie}.

In the two-dimensional $O(N)$ sigma model, 
the target space is 
the unit sphere in Euclidean $N$-dimensional space. 
The action is given by
\begin{equation}
S
=
\frac{1}{2g^2} \int d^2 x \,
\Big[
\left( \partial_i \phi^a \right)^2
+
D \left\{ \left( \phi^a \right)^2 - 1 \right\}
\Big],
\end{equation}
where $\phi^a$ with $a=1 \dots N$ are real scalar fields and 
the field $D$ is a Lagrange multiplier field 
that imposes the constraint, $\left( \phi^a \right)^2 = 1$.
The parameter $g$ is a bare coupling constant 
that needs to be renormalized.
The theory is asymptotically free, has a mass gap, and  
is therefore a good toy model for the Yang-Mills theory 
in four dimensions. 

The expectation value of the Lagrange multiplier field, 
$\left< D \right>$, serves as the mass for the $\phi$ fields.  
At large $N$, the mass gap $ \sqrt{\left< D \right>}$ can be 
computed exactly 
by looking for the saddle point of the effective potential for $D$. 
Assuming that $D$ is a constant and integrating $\phi^a$, we obtain the effective potential for $D$ as
\begin{equation}
V_{\rm eff} (D) 
=
\frac{N}{2} \left[ \int \frac{d^2p}{\left( 2 \pi \right)^2}  
\log \left( p^2  + D \right)
- \frac{D}{\lambda} \right] ,
\label{V_eff1}
\end{equation}  
where $\lambda = g^2 N$ is the 't Hooft coupling 
that is kept finite in the large-$N$ limit. 
After subtracting the UV divergence 
and renormalizing the coupling, 
the effective potential becomes 
\begin{equation}
V_{\rm eff} (D) =
-\frac{N}{8\pi} D \left( \log\frac{D}{\Lambda^2} -1 \right),
\label{V_eff2}
\end{equation}  
where the renormalization group (RG)-invariant dynamical scale 
$\Lambda$ is defined by the renormalized 't Hooft coupling 
$\lambda_\mu$ at the renormalization scale $\mu$ as 
\begin{equation}
\Lambda = \mu \exp \left( - \frac{2\pi}{\lambda_\mu} \right),
\label{Lambda}
\end{equation}
in the MS-bar scheme. 
The effective potential gives the unique minimum at 
\begin{equation}
\langle D \rangle =\Lambda^2 .
\label{minimum}
\end{equation}

Let us consider two-point correlation functions of the 
fluctuation field $\delta D(x)$ of the Lagrange multiplier field $D(x)$ 
around the expectation value $\langle D \rangle=\Lambda^2$. 
Since the correlation function is nontrivial only at the next-to-leading 
order of $1/N$ expansion, we choose a normalization 
\begin{equation}
D(x) = \Lambda^2  + \frac{\delta D (x)}{\sqrt{N}}. 
\label{eq:fluctuation}
\end{equation}
At the leading order in the large-$N$ limit, 
the two-point correlation function $\Delta(p)$ 
of the fluctuation field $\delta D(x)$ in the momentum space 
(propagator) is given as 
\begin{eqnarray} \hs{-5}
\Delta(p)
\equiv
\left[ \frac{1}{2} \int \frac{d^2q}{(2\pi)^2} 
\frac{1}{\left(q^2+ \Lambda^2 \right) \left( \left( q + p\right)^2 + \Lambda^2 \right)} \right]^{-1} =
\frac{8 \pi \sqrt{p^2 \left( p^2 + 4 \Lambda^2 \right)}}{s_p} ,
\label{CF_exact_integral}
\end{eqnarray}
where 
$s_p$ is the function of $p$ defined as 
\begin{equation}
s_p= 
4\log \left( \sqrt{\frac{p^2}{4\Lambda^2}+1} 
+ \sqrt{\frac{p^2 }{4 \Lambda^2}} \right) \hs{5}
\bigg( = 4 \, {\rm arcsinh} \frac{p}{2\Lambda} \bigg).
\label{s} 
\end{equation}

The correlation function in the position space 
can be obtained by the Fourier transformation 
\beq
\left\langle \delta D(x) \delta D(0) \right\rangle = \int \frac{d^2 p}{(2\pi)^2} e^{i p \cdot x} \Delta(p). 
\eeq
This is a well-defined UV (and IR) convergent integral. 
However, it becomes UV divergent in the limit $x=0$
\beq
\langle \delta D^2 \rangle \equiv \lim_{x \rightarrow 0} \left\langle \delta D(x) \delta D(0) \right\rangle \rightarrow \infty. 
\eeq
This quantity appears as one of the operator basis $O_n$ 
of the operator product expansion 
\begin{equation}
D(x)D(0)=\sum_{n} F_n(x) O_n, 
\label{eq:OPE}
\end{equation}
where $F_n(x)$ are the coefficient functions. 
For that reason, we are interested in the limit $x \rightarrow 0$
and call the quantity as a condensate, 
in analogy to the gluon condensate in QCD. 
To regularize the UV divergence, 
we introduce the UV cutoff $\tilde a$ 
to limit the momentum integration $|p|<\tilde a$ 
\begin{equation}
\left< \delta D^2 \right>_{\tilde a}
\equiv \int_{|p|<\tilde a} \frac{d^2p}{(2\pi)^2} \Delta(p). 
\label{eq:gluon_cond}
\end{equation}
Changing the variable from $|p|$ to $s=s_p$, 
we obtain
\beq
\left< \delta D^2 \right>_{\tilde a}
=
2 \Lambda^4 \int^{s_{\tilde a}}_{0} ds \, \frac{\cosh s-1}{s}
= 
2 \Lambda^4 \, {\rm Chin}(s_{\tilde a}),
\label{D2_ON_Exact_Real}
\eeq
where ${\rm Chin}(s_{\tilde a})$ is 
an entire function of $s_{\tilde a}$ 
related to the hyperbolic cosine integral ${\rm Chi}$
and Euler's constant $\gamma_E$ as
\beq
{\rm Chin}(s_{\tilde a}) = {\rm Chi}(s_{\tilde a}) - \log (s_{\tilde a})-\gamma_E.
\eeq
This is the regular and well-defined exact result in the large-$N$ limit \cite{Novikov:1984ac}. 
In the next section, 
instead of directly evaluating the integral \eqref{eq:gluon_cond}, 
we use the large $p/\Lambda$ expansion of the integrand 
\eqref{eq:gluon_cond}
to simulate the semiclassical expansion, 
which has an IR divergence and a renormalon type ambiguity. 
\section{Semiclassical expansion 
}

In Sec.\,\ref{sec:SemiclassicalForm}, 
we first expand the propagator  
(\ref{CF_exact_integral}) into a transseries of 
$\Lambda^2/p^2 = \exp (-4\pi/\lambda_p)$ 
and $\lambda_p$ 
in order to imitate massless perturbation theory around 
the vacuum and nontrivial backgrounds. 
We then discuss IR divergences and imaginary ambiguities in 
the expansion in Sec.\,\ref{sec:Cancellation},
and finally compute the semiclassical expansion up to order order $\Lambda^8$ in Sec.\,\ref{sec:PerturbativeExpansion}

\subsection{Expansion of the propagator in powers of $\Lambda^2/p^2$}
\label{sec:SemiclassicalForm}

Here we consider the $x\rightarrow 0$ limit 
of the correlation function, 
i.e. the condensate, 
of the fluctuation of the Lagrange multiplier field 
$\delta D(x)$ in Eq.\,\eqref{eq:gluon_cond}. 

In most of interesting theories like QCD, 
the gap equation to generate the mass gap is 
not known explicitly, 
contrary to the two-dimensional large-$N$ $O(N)$ model. 
In such a situation, we can use 
only the weak coupling perturbation theory 
with massless fields. 
We are interested in studying 
properties of perturbation theory 
and associated resurgence structure 
when only perturbative series 
with massless fields are available. 
In order to mimic such a situation, 
we use the large $p^2/\Lambda^2$ expansion 
of the propagator $\Delta(p)$ 
to obtain a transseries 
in powers of $\Lambda^2/p^2 = \exp ( - 4\pi/\lambda_p)$ 
and $\lambda_p$. 
In this way, we can study quantities such as the condensate 
as if we perform massless field perturbation theory 
on various backgrounds corresponding to 
possible nonperturbative saddle points. 
Hence we wish to expand the propagator $\Delta(p)$  in 
Eq.\,(\ref{CF_exact_integral}) in powers of $\Lambda^2/p^2$. 
The asymptotic behavior for 
$\Lambda^2 \ll p^2$ of the denominator $s_p$ 
of the propagator is given by 
\begin{equation}
s_p=4\log\left(\sqrt{\frac{p^2}{4\Lambda^2}+1}
+\sqrt{\frac{p^2}{4\Lambda^2}}\right) = \frac{8\pi}{\lambda_p} +u_p, 
\end{equation}
where the leading term is the inverse coupling $\lambda_p$ renormalized at the momentum scale $p$,\footnote{
In the large-$N$ limit, 
we do not distinguish 
the full renormalized coupling $\lambda_p$ and 
the one-loop coupling $\lambda_p'$ used in Sec.\,\ref{sec:renormalon} 
since the higher order coefficients of the beta function in \eqref{eq:beta_function} are of order $1/N$.} 
\begin{equation}
\lambda_p \equiv \frac{2\pi}{\log \left( p/\Lambda \right)}, 
\end{equation}
and the remaining term $u_p$ can be expanded in a power of $\Lambda^2/p^2$ 
\begin{equation}
u_p  =
4\log \left( \frac{1}{2}+ \sqrt{\frac{1}{4}+\frac{\Lambda^2}{p^2} } \right)=
\frac{4 \Lambda^2}{ p^2} - \frac{6 \Lambda^4}{p^4}  + \mathcal{O}(\Lambda^6) .
 \label{u_expansion}
\end{equation}
Thus, we obtain a power series expansion for 
large momenta as a power series in $u_p\lambda_p/8\pi$ 
\begin{equation}
\Delta(p)
=
p^2 \lambda_p \sqrt{1 + \frac{4\Lambda^2}{p^2}}
\sum^{\infty}_{n=0} \left(  -\frac{u_p \lambda_p}{8\pi} \right)^n,
 \label{eq:O_expansion_u}
\end{equation}
which is convergent if $u_p\lambda_p/ \left( 8\pi \right) <1 $. 
We can expand 
$\sqrt{1 + 4 \Lambda^2/p^2}$ and $u_p$ in powers 
of $\Lambda / p$ to obtain 
\begin{equation}
\Delta(p)
=
p^2 \sum^{\infty}_{l=0} \left( \frac{\Lambda}{p} \right)^{2l} 
f_{l} (\lambda_p) ,
 \label{Gamma_expansion}
\end{equation}
where $f_{l}(\lambda_p)$ is a polynomial of degree $l+1$. 
A convenient way to derive the explicit forms of $f_l(\lambda_p)$ is to use the Borel resummed form of $\Delta(p)$
\beq
\Delta(p) = 2 \pi p^2 \sum_{l=0}^\infty \left( \frac{\Lambda}{p} \right)^{2l} \int_0^\infty dt \, \left( \frac{\Lambda}{p} \right)^{t} P_l(t),
\eeq
where $P_l(t)$ is a polynomial of $t$ 
\beq
 P_l(t) \equiv \frac{(-1)^l}{l!} \Big[ (t+l+1)^{(l)} - 4 l (t+l)^{(l-1)} \Big]
 ~~~\mbox{with}~~~(a)^{(l)} = \frac{\Gamma(a+l)}{\Gamma(a)}.
\label{eq:P_l_def}
\eeq
From this expression, we find that 
$f_l(\lambda_p)$ can be obtained as
\beq
f_l(\lambda_p) = P_l(\Lambda \p_{\Lambda}) \, \lambda_p.
\label{eq:f_l_general}
\eeq

To obtain the condensate, 
we need to perform the momentum integral \eqref{eq:gluon_cond}.  
We now use the large momentum expansion \eqref{Gamma_expansion} to all momentum regions, 
including $p<\Lambda$ region. 
This is intended to imitate the calculation with massless fields, even though the large momentum expansion 
is valid only for $|p|\gg \Lambda$. 
Then we need to introduce an IR regularization, 
which is achieved by a momentum cutoff at $a$ ($|p|>a$). 
The condensate is now given as
\begin{eqnarray}
\left< \delta D^2 \right>_{\tilde a, a} \
\underset{\rm{s.c.}}{=} \
\sum^{\infty}_{l=0}  \Lambda^{2l}  C_{2l} ,
\label{eq:condensate_CF_p}
\end{eqnarray}
with
\begin{eqnarray}
C_{2l}
=
\int_{a<|p|<\tilde a} 
\frac{d^2p}{(2\pi)^2} \, p^{2-2l} \, f_{l} (\lambda_p) ,
\label{CF_p}
\end{eqnarray}
with $f_{l} (\lambda_p)$ in Eq.\,(\ref{Gamma_expansion}). 
In this work, we call the transseries expression \eqref{eq:condensate_CF_p} 
the semiclassical ansatz (s.c.), 
since this would be the transseries obtained 
through the semiclassical expansion of the path integral. 
We have introduced a UV cutoff at $\tilde a$ 
and IR cutoff at $a$ in momentum integration 
in order to eliminate the UV and IR divergences. 
However, it is not clear if the semiclassical ansatz for
$\left< \delta D^2 \right>_{\tilde a,a}$ gives 
the exact expression in the limit $a\to 0$,
since the series in powers of $\Lambda$ 
may not be convergent for $\Lambda > a$ 
and the ordering of summation and integration is exchanged. 
We will come back to this point in Sec.\,\ref{sec:Comparison}.  

Using the relation 
\beq
\frac{\lambda_p}{4\pi}  = 
\left[ \frac{4\pi}{\lambda_{\tilde a}} + \log \left(\frac{p^2}{\tilde a^2} \right) \right]^{-1}
= \sum_{n=0}^\infty \left( \frac{\lambda_{\tilde a}}{4\pi} \right)^{n+1} \left[ -\log\left(\frac{p^2}{\tilde a^2} \right) \right]^n,
\eeq 
we can expand the integrand in \eqref{CF_p} in powers of the coupling 
$\lambda_{\tilde a}$ at scale $\tilde a$ to 
\begin{eqnarray}
C_{2l}=\sum^{\infty}_{n=0} \lambda_{\tilde a}^{n+1} c_{(2l,n)} , 
\label{eq:trnasseries_mu_2l}
\end{eqnarray}
whose explicit computations 
for $l=0,\dots, 4$ are given in the Appendix \ref{sec:perturbation}. 

The power series in Eq.\,\eqref{eq:trnasseries_mu_2l} 
can contain factorially divergent parts, 
which have a precise meaning by the Borel resummation. 
If such divergent series are Borel non-summable,
the associated imaginary ambiguities should
be of the renormalon type, 
since only renormalon type ambiguities are 
expected to remain in the large-$N$ limit. 
The $l=0$ terms $c_{(0,n)}$ correspond to the usual 
perturbative expansion on the trivial vacuum. 
The physical interpretation of $c_{(2l,n)}$ for higher $l>0$ is 
that it is a contribution of the fluctuation 
at order $\lambda_{\tilde a}^{n+1}$ 
around a possible semi-classical configuration 
$(\Lambda/\tilde a)^{2l} \sim e^{-4 \pi l / \lambda_{\tilde a}}$, 
although we have no understanding of 
such a semiclassical configuration explicitly.  

\subsection{Infrared divergence and imaginary ambiguities}
\label{sec:Cancellation}

It is evident that there are three issues with the semiclassical 
expansion obtained above due to the IR behavior. 
The first one is that the integral $C_{2l}$ is IR divergent when 
$l \geq 2$ due to the factor $p^{2-2l}$ in the integrand, 
which requires an IR cutoff $a$. 
We need to take the limit $a\to 0$ at 
the end of the calculation. 
The second issue then arises 
when the IR cutoff is small $a \ll \Lambda$, 
because the semiclassical ansatz above
involves a power series in $\Lambda^2/a^2$ 
and requires a care to take the limit $a\to 0$. 
We will come back to this point in Sec.\,\ref{sec:Comparison}.
The third issue is that there is a possible singularity
at $p=\Lambda$ due to the terms 
involving the renormalized coupling constant 
$\lambda_p = 4\pi/\log (p^2/\Lambda^2)$ 
in Eq.\,\eqref{CF_p}. 
In fact, the renormalon ambiguity 
in the usual perturbation theory 
is due to this type of singularity in the integrand of $C_{0}$.  
Below we identify these singularities 
in the integrand of all $C_{2l}$. 
Changing variables from $p$ to 
$\tilde t = \log (\tilde a^2 /p^2) = 4\pi/\lambda_{\tilde a} - 4\pi/\lambda_p$, 
we can rewrite it as
\begin{equation}
C_{2l}
= \frac{1}{4 \pi}
\int^{\log(\tilde a^2/a^2)}_{0} d \tilde t \, \left( \tilde a^2 e^{- \tilde t } \right)^{2-l}
f_{l} \left(\frac{4\pi}{4 \pi / \lambda_{\tilde a}-\tilde t} \right).
\label{deltaD_t}
\end{equation}
This form resembles the Borel resummation of a divergent 
perturbative series. 
For $l <2$ we can take $a \rightarrow 0$ at this point, and 
$C_{2l}$ becomes a Borel resummation. 
For $l \geq 2$, we cannot take $a \rightarrow 0$ 
due to the IR divergence. 
Since $f_l(\lambda_p)$ is a polynomial of order $l+1$ and hence 
the integrand has a pole at $\tilde t = 4 \pi/ \lambda_{\tilde a}$.
If $a<\Lambda$, 
this pole is on the integration contour of \eqref{deltaD_t} 
since $0 < 4 \pi / \lambda_{\tilde a} <\log (\tilde a^2 /a^2)$. 

In order to circumvent the poles, 
we use an analytic continuation 
of the coupling $\lambda_{\tilde a}$ to the complex plane. 
After the integration over $t$, 
we then analytically continue back to 
the real axis in two different directions: 
\begin{equation}
\lambda_{\tilde a} \rightarrow \lambda_{\tilde a} \pm i \epsilon ,
\label{AC_lambda}
\end{equation}
with $\epsilon>0$, or equivalently 
$\Lambda \rightarrow \Lambda \left( 1 \pm i \epsilon' \right)$ 
with $\epsilon' = 2 \pi \epsilon / \lambda_{\tilde a}^2 >0$. 
We then take $\epsilon$ to zero in the end. 
This can be understood as a deformation of the integration contour 
in Eq.\,(\ref{deltaD_t}) in the upper or lower $t$-plane.  

The deformation of the integration contour 
can give rise to an ambiguity, 
since the imaginary part of $C_{2l}$ depends 
on whether we take $\lambda_{\tilde a} + i \epsilon $ 
or $\lambda_{\tilde a} - i \epsilon$.  
The imaginary ambiguities, however,
 should cancel once we sum over all $l$, 
 regardless of the prescription. 
We can find the imaginary ambiguities 
by computing the residue.  
Using Eq.\,(\ref{deltaD_t}) and computing up to order $\Lambda^8$, 
we find that our semiclassical ansatz \eqref{eq:condensate_CF_p}
as a whole is indeed free of imaginary ambiguity:
\begin{equation}
{\rm{Im}} \left< \delta D^2 \right>_{\tilde a, a} \ \underset{\rm s.c.}{=} \ 
\pm \pi 
\left[
\left( \tilde{a}^2 e^{-\frac{4\pi}{\lambda_{\tilde a}}} \right)^2 \Lambda^0 
-
2 \Lambda^4
+
\left(\tilde{a}^2 e^{-\frac{4\pi}{\lambda_{\tilde a}}} \right)^{-2} \Lambda^8
\right] \theta(\Lambda -a )
=
0 .
\label{Delta_deltaD}
\end{equation}
We show that only the three terms, $C_0$, $C_4$, and $C_8$, have non-zero residues at $t=4\pi/\lambda_{\tilde a}$ 
that give rise to 
the imaginary ambiguities in Sec.\,\ref{sec:PerturbativeExpansion}.
We also show that the first term at order 
$\Lambda^0$ in the bracket corresponds to the renormalon 
ambiguity due to the Borel resummation of the divergent perturbative 
series on the trivial vacuum.  
Thus, the ansatz \eqref{eq:condensate_CF_p} gives a surprising 
result that the renormalon ambiguity on the trivial vacuum 
(order $\Lambda^0$) is cancelled not solely by the ambiguity 
from the term at order $\Lambda^4$ as one would naively expect, 
but the combination of the terms at order $\Lambda^4$ and $\Lambda^8$.

\subsection{Perturbative expansion around 
vacuum and nontrivial background}
\label{sec:PerturbativeExpansion}

In this section, we compute 
the coefficients $C_{2l}$ of the expansion \eqref{eq:condensate_CF_p} 
and investigate the origin of each ambiguity 
in Eq.\,(\ref{Delta_deltaD}).  
We first take a large IR cutoff, $\Lambda \ll a < \tilde a$, where 
the expansion in powers of $\Lambda^2/p^2~(a<p<\tilde a)$ 
of the integrand is convergent and well-defined.  
This allows us to obtain unambiguous $C_{2l}$ 
without any imaginary parts. 
We then take a small cutoff $a < \Lambda$. 
As explained in the previous section, 
we use an analytic continuation of 
$\lambda_{\tilde a}$ (or $\Lambda$) as \eqref{AC_lambda}
to avoid a possible singularity at $p = \Lambda$. 
Depending on the sign of $ \pm i \epsilon $, 
we show that $C_{2l}$ 
picks up an imaginary part in accordance 
with Eq.\,(\ref{Delta_deltaD}). 

The integral for $C_{2l}$ gives
\begin{eqnarray}
C_{2l} 
=
\int^{\tilde a}_a \frac{dp}{2\pi} \, p^{3-2l} \, 
f_{l} (\lambda_p)
= \left. \mathcal C_{2l}(p)\right|_a^{\tilde a} 
= \mathcal C_{2l}(\tilde a) - \mathcal C_{2l}(a) ,
\label{C_2l_v}
\end{eqnarray}
where we have defined $\mathcal C_{2l}(p)$
as an indefinite integral of the $p$-integration. 
We call $\mathcal C_{2l}(\tilde a)$ and $\mathcal C_{2l}(a)$ 
as the UV and IR contributions, respectively, 
although only the difference is unambiguously defined. 


We now compute $C_{2l}$ for $l=0, \dots, 4$.  
In the  semiclassical expansion, 
one would first need to compute 
the coefficients $c_{2l}$ of perturbative expansion, 
and then (Borel) resum it to obtain 
$C_{2l} = \sum^{\infty}_{n=0} \lambda_{\tilde a}^{n+1} c_{\left( 2l, n \right)}$. 
We demonstrate this for the case of $l=0$ here, 
and the rest in Appendix \ref{sec:perturbation}.  
Alternatively we can directly compute $C_{2l}$ 
from Eq.\,(\ref{CF_p}). 

The leading contribution, the term at order $\Lambda^{0}$, is given as
\begin{equation}
c_{\left( 0, n \right)}
=
\int_{a<|p|<\tilde a}\frac{d^2p}{(2\pi)^2} \, p^{2}
\left( \frac{1}{4\pi} \log\frac{\tilde a^2}{p^2} \right)^n 
=
\frac{\tilde a^4}{\left( 8 \pi \right)^{n+1} } 
\left[ \Gamma(n+1) - \Gamma\left(n+1, 2 \log \frac{\tilde a^2}{a^2} \right) \right],
\end{equation}
where $\Gamma(n+1,\alpha)$ is the incomplete Gamma function
\beq
\Gamma(n+1,\alpha) = \int_{\alpha}^\infty dt \, e^{-t} t^n. 
\eeq
If we turn off the IR cutoff $a \rightarrow 0$, 
the second term vanishes, 
and we arrive at the known perturbative result. 
If we keep an arbitrary IR cutoff $a$, 
then we have $C_0 = \mathcal C_0(\tilde a) - \mathcal C_0(a)$ with 
\begin{eqnarray}
\mathcal C_0 (p)
&=&
\tilde a^4
\sum^{\infty}_{n=0} 
\left( \frac{\lambda_{\tilde a}}{8\pi} \right)^{n+1} 
\Gamma \left( n+1, 2 \log \frac{\tilde a^2}{p^2} \right).
\end{eqnarray}
This is a divergent asymptotic series since $\Gamma(n+1,\alpha)\sim n!$ for large $n$. 
Applying the Borel resummation, we obtain 
\beq
\mathcal C_0 (p) =
-p^4 \int^{\infty}_{0} dt \,
\frac{e^{-t}}{t - \frac{8\pi}{\lambda_p}}
= p^4 e^{-8\pi/\lambda_p} \, \left[ \gamma_E + \log \left( - \frac{8\pi}{\lambda_p} \right) - {\rm Ein} \left( - \frac{8\pi}{\lambda_p} \right) \right],
\label{eq:C0_borel}
\eeq
where ${\rm Ein}(z)$ denotes the entire function defined as\footnote{The standard exponential integral ${\rm Ei}(z)$
is related to the entire function ${\rm Ein}(z)$ as
\beq
{\rm Ei}(z) = \gamma_E + \log z - {\rm Ein}(-z),  
\eeq
}
\beq
{\rm Ein}(z) = \int_0^z dt \, \frac{1-e^{-t}}{t}. 
\eeq 
Due to the branch cut of 
$\log(-8\pi/\lambda_p) = \log(-2 \log p^2/\Lambda^2)$, 
the function $\mathcal C_0 (p)$ is ambiguous for $p>\Lambda$
\beq
{\rm Im} \, \mathcal C_0 (p) 
\, = \, \pm \pi p^4 \exp \left( -\frac{8\pi}{\lambda_p} \right) \, \theta(p-\Lambda) 
\, = \, \pm \pi \Lambda^4 \, \theta(p-\Lambda).
\eeq
The total imaginary ambiguity at the leading order can be then 
expressed as
\begin{equation}
\mathrm{Im} \, C_0 
\, = \, \mathrm{Im} \, \mathcal C_0(\tilde a) -  \mathrm{Im} \, \mathcal C_0(a) 
\, = \,
\pm \left\{ \pi -  \pi \theta(a - \Lambda) \right\} \Lambda^4    
\, =\,
\pm \pi \Lambda^4 \, \theta( \Lambda - a ),
\label{eq:ambiguity_C0}
\end{equation}
where we have assumed that the UV scale $\tilde a$ is always larger than $\Lambda$. 
While there is a usual renormalon ambiguity when $a<\Lambda$, 
the imaginary ambiguity is absent when $a>\Lambda$.

The $\Lambda^2$ and $\Lambda^6$ can be readily computed without any imaginary ambiguities. 
For notational simplicity, we use $v_p$ defined as
\beq
v_p \equiv \frac{4\pi}{\lambda_p} = \log \frac{p^2}{\Lambda^2},
\eeq
instead of $\lambda_p$. 
At order $\Lambda^2$, we have
\begin{equation}
\mathcal C_2(p)
=
\int dp \frac{4p \left(-1 +  v_p \right)}{v^2_p}
=
\frac{2 p^2 }{v_p},
\end{equation}
while at $\Lambda^6$, we obtain
\begin{equation}
\mathcal C_{6}(p)
=
\int dp \frac{-48-24 v_p + 20 v^2_p +24 v^3_p}{3 p^3 v^4_p}
=
\frac{8+ 2 v_p -12 v^2_p}{3 p^2 v^3_p}.
\end{equation}
We thus find that the IR contribution 
$\mathcal C_2(a)$ goes to zero, 
while $\mathcal C_6(a)$ diverges as $a$ goes to zero. 
Note that each of the integrands for $\mathcal C_2(p)$ and $\mathcal C_6(p)$ has a pole at $p = \Lambda$ 
but the residue is zero and hence it does not give any ambiguities.

At order $\Lambda^4$, we have
\begin{equation}
\mathcal C_4 (p)
= \int dp \frac{8 - 2 v_p -4 v^2_p }{p v^3_p}
= - 2 \log \left(v_p \right) - \frac{2 - v_p}{v^2_p}.
\end{equation}
This term is also IR divergent 
$\mathcal C_4(a) \rightarrow \infty ~(a \rightarrow 0)$.  
Moreover the logarithm gives rise 
to the imaginary ambiguity when $v_p<0 $: 
\begin{equation}
\mathrm{Im} \, C_4 =  \mathrm{Im} \, \mathcal C_4(\tilde a) -  \mathrm{Im} \, \mathcal C_4(a)
= 
\mp 2 \pi  \theta(\Lambda-a).
\end{equation}
Compared to the renormalon ambiguity \eqref{eq:ambiguity_C0}, this ambiguity at order 
$\Lambda^4$ has the opposite sign but its magnitude is twice as 
large, so the renormalon ambiguity is not cancelled if we stop 
the calculation at this order.  

At order $\Lambda^8$, we have 
\beq
\mathcal C_8(p) &=&
\int dp \frac{96 + 120 v_p + 22 v^2_p - 59 v^3_p - 60 v^4_p}{3 p^5 v^5_p} \notag \\
&=&
\frac{1}{\Lambda^4} \left[  -{\rm Ein} \left( \frac{8\pi}{\lambda_p} \right) + \log \left( \frac{8\pi}{\lambda_p} \right) + \gamma_E \right]
-\frac{24 + 24 v_p -13 v^2_p - 33 v^3_p}{6 p^4 v^4_p}.
\label{C8_p}
\eeq
At this order, 
the logarithm remains as in the case of $C_{0}(p)$. 
Therefore it has the imaginary ambiguity when 
$v_a <0$ or $a < \Lambda$:
\begin{equation}
\mathrm{Im} \, C_8
= \pm \theta (\Lambda-a) \frac{\pi }{\Lambda^4}.
\end{equation}
Using Eq.\,(\ref{c8_p}) in Appendix, we can write the perturbative expansion as
\begin{equation}
\mathcal C_8 (p)
\supset
\frac{1}{{\tilde a}^4} \sum^{\infty}_{n=0} \left( - \frac{\lambda_{\tilde a}}{8 \pi} \right)^{n+1} \Gamma \left(n+1, -2 \log \frac{\tilde a^2}{p^2} \right)
=
-\frac{1}{p^4} \int^{\infty}_{0} dt \, \frac{e^{-t}}{t + \frac{8\pi}{\lambda_p}}.
\label{eq:borel_mu}
\end{equation}
The integrand has a pole 
at $t=-8\pi/\lambda_p$ and 
the residue gives the imaginary ambiguity of Eq.\,(\ref{C8_p}). 
One should note that the $t$-plane pole 
for $\mathcal C_{2l}(p)$ is at $t=-8\pi/\lambda_p$, in contrast to 
$t=8\pi/\lambda_p$ for $\mathcal C_0(p)$ in \eqref{eq:C0_borel}.

By using Eq.\,\eqref{eq:f_l_general}, 
we can show that $\mathcal C_{2l}(p)$ for general $l$ is given by
\beq
\mathcal C_{2l}(p) = p^{4-2l} \int_0^\infty dt \left( \frac{\Lambda}{p} \right)^t \frac{P_l(t)}{4-2l-t} = - P_l(\Lambda \p_\Lambda) \left[ \Lambda^{-2l+4} \,\Gamma \left( 0, (l-2) \log \frac{p^2}{\Lambda^2} \right) \right]. 
\label{eq:C_l_general}
\eeq
From this expression, we can check that there is no ambiguity for $l \geq 5$.

We now combine all the results up to order $\Lambda^8$ obtained above and write the semi-classical expansion of the condensate for any values of $\tilde{a} > \Lambda$ and $a \neq \Lambda$
\begin{eqnarray}
\left< \delta D^2 \right>_{\tilde a, a} &\underset{\rm s.c.}{=}& \
\sum^{\infty}_{l=0} \Lambda^{2l} \Big[ \left\{ \mathcal C_{2l} (\tilde a) \right\} -
\left\{ \mathcal C_{2l} (a) \right\} \Big] \label{CF_SC_Real}
 \\
&=& \phantom{+}
\Lambda^0
\left[ 
\tilde a^4
\left\{ e^{-8\pi/\lambda_{\tilde a}} {\rm Ei} \left( \frac{8\pi}{\lambda_{\tilde a}} \right) \right\} 
- a^4 \left\{ 
e^{-8\pi/\lambda_{a}} {\rm Ei} \left( \frac{8\pi}{\lambda_{a}} \right) \right\} 
\pm i\pi\Lambda^4 \theta(\Lambda -a )
\right]
\nonumber \\
&& +
\Lambda^2
\left[ 
{\tilde a}^2 
\left\{ \frac{\lambda_{\tilde a}}{2 \pi} \right\} - 
a^2 \left\{ \frac{\lambda_a}{2 \pi} \right\}
\right]
\nonumber
\\
&&
+
\Lambda^4
\left[ 
{\tilde a}^0
\left\{ 
\frac{\lambda_{\tilde a}}{4 \pi} - \frac{\lambda_{\tilde a}^2}{8 \pi^2} - 2 \log \left( \frac{4 \pi}{\lambda_{\tilde a}} \right) 
\right\}
- a^0 \left\{ 
\frac{\lambda_a}{4 \pi} - \frac{\lambda_a^2}{8 \pi^2} - 2 \log  \left| \frac{4 \pi}{\lambda_a} \right| 
\right\}
\mp 2 \pi i \theta (\Lambda - a)
\right]
\nonumber
\\
&&
+
\Lambda^6   
\left[ 
\frac{1}{\tilde a^2}
\left\{
-\frac{\lambda_{\tilde a}}{\pi} + \frac{\lambda_{\tilde a}^2}{24\pi^2} + \frac{\lambda_{\tilde a}^3}{24 \pi^3} \right\}
- \frac{1}{a^2} \left\{
-\frac{\lambda_a}{\pi} + \frac{\lambda_a^2}{24\pi^2} + \frac{\lambda_a^3}{24 \pi^3} \right\} \right]
\nonumber
\\
&&
+
\Lambda^8 \left[ 
\frac{1}{\tilde a^4} 
\left\{
e^{8\pi/\lambda_{\tilde a}} 
\mathrm{Ei} \left(- \frac{8 \pi }{ \lambda_{\tilde a}} \right)
+
\frac{11 \lambda_{\tilde a}}{8\pi} + \frac{13 \lambda_{\tilde a}^2}{96 \pi^2} 
-\frac{\lambda_{\tilde a}^3}{16 \pi^3} - \frac{\lambda_{\tilde a}^4}{64 \pi^4} 
\right\}
\right.
\nonumber
\\
&&
~~~~ \left.
- \frac{1}{a^4} \left\{
e^{8\pi/\lambda_a} 
\mathrm{Ei} \left(- \frac{8 \pi }{ \lambda_a} \right)
+
\frac{11 \lambda_a}{8\pi} + \frac{13 \lambda_a^2}{96 \pi^2} 
-\frac{\lambda_a^3}{16 \pi^3} - \frac{\lambda_a^4}{64 \pi^4} 
\right\}
\pm \frac{i \pi  }{\Lambda^4} \theta (\Lambda-a) 
\right] \nonumber \\
&& +\mathcal{O} (\Lambda^{10}), 
\phantom{\bigg[}
\nonumber
\end{eqnarray}
where the exponential integral ${\rm Ei}(z)$ is defined as
\beq
{\rm Ei}(z) = \gamma_E + \log z - {\rm Ein}(-z) = \gamma_E + \log z - \int_0^{-z} dt \, \frac{1-e^{-t}}{t}.
\eeq 
The imaginary ambiguity at each order 
depends on the value of the IR cutoff $a$
as denoted by the Heaviside step function $\theta(\Lambda-a)$.
For a large IR cutoff $a>\Lambda$, 
there is no ambiguity at any order. 
Once we take the small cutoff $a <\Lambda$, 
imaginary ambiguities appear at order 
$\Lambda^0$, $\Lambda^4$, and $\Lambda^8$. 
We have identified the imaginary ambiguity 
at order $\Lambda^0$ as 
the renormalon ambiguity in perturbation theory 
on the trivial vacuum.  
The imaginary ambiguities 
at order $\Lambda^4$ and $\Lambda^8$ 
also arise when $a<\Lambda$, 
and the combination of the two cancels 
the renormalon ambiguity, 
leaving the semiclassical expansion 
free of imaginary ambiguities as a whole. 
This result agrees with Eq.\,(\ref{Delta_deltaD}).

Using the general form $C_{2l}$ in Eq.\,\eqref{eq:C_l_general}, 
the all-order transseries can be written as
\beq
\langle \delta D^2 \rangle_{\tilde a,a} \ \underset{\rm s.c.}{=} \ \mu^4 \sum_{l=0}^\infty \left(\frac{\Lambda}{\mu} \right)^{2l} \int_0^\infty dt \, \left( \frac{\Lambda}{\mu} \right)^t \mathcal B_l(t).
\label{eq:condensate_Borel}
\eeq
The Borel transform $\mathcal B_l(t)$ is given by
\beq
\mathcal B_l(t) = \frac{1}{2} \mu^{-2\eta_l(t)} \Big[ {\tilde a}^{2\eta_l(t)} - a^{2\eta_l(t)} \Big] \frac{P_{l}(t)}{\eta_l(t)}, 
~~~ \mbox{with} ~~~ \eta_l(t) = 2 - l - \frac{t}{2},
\eeq
where $P_l(t)$ is the polynomial given in Eq.\,\eqref{eq:P_l_def}.
Since the Borel transform $\mathcal B_l(t)$ 
has no pole on the positive real axis, 
there is not ambiguity in this expression. 
However, the integral converges only when $a > \Lambda$. 
For $a < \Lambda~(\lambda_a < 0)$, 
the Borel resummation for the $a$-dependent term must be performed along the negative real axis\footnote{
The integration path of the Borel resummation must be chosen depending on the sign (or, more precisely, argument) of the variable as 
\beq
\sum_{n=0}^\infty a_n \lambda^n = 
\begin{cases} 
+ \displaystyle \int_0^{+\infty} dt \, e^{-t/\lambda} B(t) & \mbox{for $\lambda>0$} \\
- \displaystyle \int_{-\infty}^0 dt \, e^{-t/\lambda} B(t) & \mbox{for $\lambda<0$}
\end{cases} ~~~\mbox{with}~~~ B(t) = \sum_{n=0}^\infty \frac{a_n}{\Gamma(n)} t^{n-1}.
\eeq
}, or equivalently, 
the Borel resummation must be rewritten as
\beq
\langle \delta D^2 \rangle_{\tilde a,a} \ \underset{\rm s.c.}{=} \ \mu^4 \sum_{l=0}^\infty \left(\frac{\Lambda}{\mu} \right)^{2l} \int_{-\infty}^\infty dt \, \left( \frac{\Lambda}{\mu} \right)^t \tilde{\mathcal B}_l(t),
\eeq
with
\beq
\tilde{\mathcal B}_l(t) = \frac{1}{2} \mu^{-2\eta_l(t)} \Big[ {\tilde a}^{2\eta_l(t)} \theta(t) + a^{2\eta_l(t)} \theta(-t) \Big] \frac{P_{l}(t)}{\eta_l(t)},
\eeq
where $\theta(t)$ is the step function. 
In this case, $\tilde{\mathcal B}_l(t)$ with $l=0,2,4$ have singularities at $t=4,0,-4$ and give rise to the imaginary ambiguities at order $\Lambda^0$, $\Lambda^4$ and $\Lambda^8$, respectively. 
It is worth noting that the singularity 
on the negative real axis on the Borel plane 
is relevant when $a < \Lambda$. 
This is related to the fact that the condensate contains 
terms with the negative coupling constant 
$\lambda_a = 2\pi/\log(a/\Lambda)$ 
and the non-perturbative factors $(\Lambda/a)^{2l}$ 
that become more dominant for higher $l$.
This is a typical situation in which 
renormalons give rise to imaginary ambiguities.

Although the condensate in the semiclassical expansion is real 
for any value of the IR cutoff $a$, the expansion is convergent 
only if the IR cutoff is large, $a \gg \Lambda$. When the cutoff 
is small,  $a \ll  \Lambda$, the terms in the series for the IR 
contribution, $C_l (a)$, becomes divergent for $l \geq 2$. 
This tells us that we need to consider taking $a \gg \Lambda$ 
in order to sum over $l$. We can take the limit $a\to 0$ 
only after summing over $l$. 
We now discuss this procedure 
in the next section.

\section{Transseries from the exact result 
}
\label{sec:Comparison}

We reanalyze the exact result 
for the condensate $\langle \delta D^2 \rangle$ 
in Eq.\,\eqref{D2_ON_Exact_Real}, 
in order to understand 
the newly found imaginary ambiguities 
at higher powers of $\Lambda$ 
and to take the $a\to 0$ limit properly 
in our result in Eq.\,\eqref{CF_SC_Real}. 
To compare the exact result with 
the semiclassical expansion Eq.\,\eqref{CF_SC_Real}, 
we now work out the transseries representation of 
the exact result in Eq.\,\eqref{D2_ON_Exact_Real}. 
From dimensional reasons, 
the condensate is a function of a single variable 
$\Lambda/{\tilde a}$ apart from the factor $\Lambda^4$ 
\begin{equation}
\left< \delta D^2 \right>_{\tilde a}
=\Lambda^4 F\left(s_{\tilde a}\right)
= 
2 \Lambda^4 \int^{s_{\tilde a}}_{0} ds \, \frac{\cosh s-1}{s},
\label{eq:def_F}
\end{equation}
where the upper end $s_{\tilde a}$ of the 
integral is a function of $\Lambda^2/{\tilde a}^2$ 
as defined in Eq.\,\eqref{eq:O_expansion_u}
\begin{equation}
s_{\tilde a}
=\frac{8\pi}{\lambda_{\tilde a}}+u_{\tilde a}, 
\quad 
u_{\tilde a}=4\log\left(\frac{1}{2}+\sqrt{\frac{1}{4}
+\frac{\Lambda^2}{{\tilde a}^2}}\right) .
\label{eq:s_mu_u_mu}
\end{equation}
As given in Eq.\,\eqref{eq:u_mu_power_series} 
in Appendix \ref{sec:power_series}, 
the variable $u_{\tilde a}$ can be expanded in powers 
of $\Lambda^2/{\tilde a}^2$ with the finite radius of convergence. 
On the other hand, the function $F(s_{\tilde a})$ 
can be expanded in powers of $u_{\tilde a}$ 
with a finite radius of convergence. 
Therefore, we find that contributions 
from the integration region $8\pi/\lambda_{\tilde a} < s < s_{\tilde a}$ 
in the integral representation in 
Eq.\,\eqref{eq:def_F} of 
the exact solution gives a power series 
in $\Lambda^2/{\tilde a}^2$. 
Moreover, it is easy to see that each $l$-th order terms 
$\Lambda^{2l}/{\tilde a}^{2l}$ contains only up to $l$ powers of 
$\lambda_{\tilde a}$. 
The remaining term, however, gives a divergent power series 
in $\lambda_{\tilde a}$ and needs to be Borel resummed. 
In fact, the contribution from 
$0<s<8\pi/\lambda_{\tilde a}$ can be rewritten into 
the Borel resummation of the factorially divergent series 
\begin{eqnarray}
&& F\left(\frac{8\pi}{\lambda_{\tilde a}}\right)
=
- \frac{{\tilde a}^4}{\Lambda^4} \int_0^\infty dt \frac{e^{-t}}{t-\frac{8\pi}{\lambda_{\tilde a}}\pm i 0}
+ \left[ 2\log\left(\frac{\lambda_{\tilde a}}{8\pi} \right)
-2\gamma_{\rm E}\mp i\pi 
\right]
-\frac{\Lambda^4}{{\tilde a}^4} \int_0^\infty dt \frac{e^{-t}}
{t+\frac{8\pi}{\lambda_{\tilde a}}}
.
\label{eq:Borel_resum_mu}
\end{eqnarray}
We note that the first term is 
the result of Borel resummation of 
Borel-nonsummable divergent power series and 
has an imaginary ambiguity, 
which is cancelled by the imaginary 
ambiguity in the second term \cite{Novikov:1984ac}. 
The third term is the result of Borel resummation 
of Borel-summable divergent series 
without imaginary ambiguity. 
Combining contributions from the integration region 
$8\pi/\lambda_{\tilde a}<s<s_{\tilde a}$, 
we obtain up to terms of order $\Lambda^8$ as 
\begin{eqnarray}
\left< \delta D^2 \right>_{\tilde a}
&=&
{\Lambda^0}{{\tilde a}^4} \left\{- \int_0^\infty dt 
\frac{e^{-t}}{t-\frac{8\pi}{\lambda_{\tilde a}}\pm i 0} \right\}
+{\Lambda^2} {{\tilde a}^2} \left\{\frac{\lambda_{\tilde a}}{2 \pi}\right\}  
\nonumber
\\
&&
+
{\Lambda^4} \left\{ 
\frac{\lambda_{\tilde a}}{4 \pi} - \frac{\lambda^2_{\tilde a}}{8 \pi^2}
+2 \log \left( \frac{\lambda_{\tilde a}}{8 \pi} \right) -2\gamma_{\rm E}
\mp i\pi \right\} 
+
\frac{\Lambda^6}{{\tilde a}^2}
\left\{ -\frac{\lambda_{\tilde a}}{\pi} + \frac{\lambda_{\tilde a}^2}{24\pi^2} 
+ \frac{\lambda_{\tilde a}^3}{24 \pi^3}  \right\}
\nonumber
\\
&&
+
\frac{\Lambda^8}{{\tilde a}^4}
\left\{
- \int_0^\infty dt \frac{e^{-t}}{t+\frac{8\pi}{\lambda_{\tilde a}}}
+ \frac{11 \lambda_{\tilde a}}{8\pi} + \frac{13 \lambda^2_{\tilde a}}{96 \pi^2} -\frac{\lambda_{\tilde a}^3}{16 \pi^3} - \frac{\lambda_{\tilde a}^4}{64 \pi^4} 
\right\}
+\mathcal{O} \left(\frac{\Lambda^{10}}{{\tilde a}^{6}}\right)  .
\label{eq:trasseries_condensate} 
\end{eqnarray}
This is the Borel resummed transseries
for the exact result without an IR cutoff, 
which is valid for ${\tilde a} \gg \Lambda$. 
We can see that the ambiguity structure
of this transseries without the IR cutoff 
is different from that with the IR cutoff \eqref{CF_SC_Real}.

Although the condensate itself has no IR divergence, 
we can introduce the IR cutoff $a$ 
for the momentum integration 
in order to compare the result of 
the semiclassical ansatz with the exact result 
\begin{equation}
\left< \delta D^2 \right>_{\tilde a,a}
=\Lambda^4 \left\{F\left(s_{\tilde a}\right)- F\left(s_{a}\right)\right\}.
\label{eq:def_F2}
\end{equation}
The contribution $F(s_a)$ is defined by the integral 
representation in Eq.\,\eqref{eq:def_F}, 
with the upper end of integration given 
by $s_a$ instead of $s_{\tilde a}$. 
We find that it is expandable in power series of $a/\Lambda$ 
as given in Eq.\,\eqref{eq:integral_rep_Fa} in Appendix \ref{sec:power_series}. 
In particular, $F(s_a)\to 0$ in the limit of $a\to 0$:
\begin{equation}
F(s_a) = \frac{a}{\Lambda} + \mathcal{O} \left(\frac{a^{2}}{\Lambda^{2}}\right) . 
\end{equation}
On the other hand, the function $F(s_a)$ 
has an interesting analytic structure. 
It has a Borel resummed transseries form for 
$a \gg \Lambda$ whose functional form is precisely 
identical to that in Eq.\,\eqref{eq:trasseries_condensate}. 
In this region, the Borel non-summable divergent series 
in $\lambda_{\tilde a}>0$ gives imaginary ambiguities 
which cancel those from the contribution $F(s_{\tilde a})$. 

To understand the result in Eq.\,\eqref{CF_SC_Real} of the 
semiclassical ansatz, let us first consider the Borel resummed 
transseries valid for $a \gg \Lambda$. 
It consists of a series in powers of $\Lambda^2/a^2$, whose $l$-th 
power coefficient is a (divergent) power series of $\lambda_a$, 
in exactly the same form as that in Eq.\,\eqref{eq:trasseries_condensate} with 
$a$ replacing $\tilde{a}$. 
If we take the coefficient of each term of $(\Lambda/a)^{2l}$ 
and analytically continue each coefficient to the region $a<\Lambda$, 
we find the following formal expression similar to a transseries 
\begin{eqnarray}
\Lambda^4F(s_a)_{\rm formal}
&=&
{\Lambda^0}{a^4}\left\{- \int_0^\infty dt 
\frac{e^{-t}}{t-\frac{8\pi}{\lambda_a}}\right\}
+{\Lambda^2} {a^2} \left\{\frac{\lambda_a}{2 \pi}\right\}  
\label{eq:formal_trasseries} 
\\
&&
+
{\Lambda^4}\left\{ 
\frac{\lambda_a}{4 \pi} - \frac{\lambda_a^2}{8 \pi^2}
+2 \log \left( \frac{-\lambda_a}{8 \pi}\right) - 2\gamma_{\rm E}
\pm i \pi \right\} 
+
\frac{\Lambda^6}{a^2}
\left\{ -\frac{\lambda_a}{\pi} + \frac{\lambda_a^2}{24\pi^2} 
+ \frac{\lambda_a^3}{24 \pi^3}  \right\}
\nonumber
\\
&&
+
\frac{\Lambda^8}{a^4}
  \left\{
- \int_0^\infty dt \frac{e^{-t}}
{t+\frac{8\pi}{\lambda_a}\mp i 0}
+
\frac{11 \lambda_a}{8\pi} + \frac{13 \lambda_a^2}{96 \pi^2} 
-\frac{\lambda_a^3}{16 \pi^3} - \frac{\lambda_a^4}{64 \pi^4} 
\right\}
+\mathcal{O} \left(\frac{\Lambda^{10}}{a^{6}}\right) .
\nonumber
\end{eqnarray}
Since $\lambda_a=4\pi/\log(a^2/\Lambda^2)<0$ for $a<\Lambda$, 
the $\Lambda^0$ term becomes Borel summable, whereas the $\Lambda^8$ 
term becomes Borel nonsummable, resulting in an imaginary ambiguity. 
We also need an analytic continuation for the $\Lambda^4$ term. 
Thus this formal transseries exhibits imaginary ambiguities 
in the $\Lambda^4$ and $\Lambda^8$ terms. 
We now observe that the result of the semiclassical ansatz 
in Eq.\,\eqref{CF_SC_Real} is 
precisely recovered as the difference of 
$F(s_{\tilde{a}})$ in Eq.\,\eqref{eq:trasseries_condensate} 
and this formal transseries $F(s_a)$ 
in Eq.\,\eqref{eq:formal_trasseries}.
In the semiclassical ansatz, we note that only the difference 
between the UV and IR contributions is determined. 

Now we can understand the imaginary ambiguities found for 
$a<\Lambda$ in Eq.\,\eqref{CF_SC_Real} using the semiclassical 
ansatz in Eq.\,\eqref{CF_p}. 
In the semiclassical ansatz, we first expand the momentum integrand 
in powers of $\Lambda^2/p^2$ which is valid only for $p^2 \gg \Lambda^2$. 
We then evaluate the momentum integral of each powers of 
$\Lambda^{2}/p^2$ using an IR cutoff $|p|>a$. 
As a result, the IR contribution $\mathcal C_{2l}(a)$ 
for the $\Lambda^{2l}$ 
term involves powers of $(\Lambda/a)^{2l}$. 
However, we are using the expansion 
in powers of $\Lambda^2/p^2$ 
outside of its validity, when we take the IR cutoff $a$ smaller 
than the dynamical mass $\Lambda$. 
This is the reason why we obtain the imaginary ambiguity corresponding 
to the Borel non-summable series in $\lambda(a)$ at order $\Lambda^8/a^4$ 
in Eq.\,\eqref{eq:formal_trasseries} of the formal transseries 
$F(s_a)_{\rm formal}$. 

In order to properly take the limit of $a\to 0$ of the result 
in Eq.\,\eqref{CF_SC_Real} of the semiclassical ansatz, 
we need to first continue $a$ 
from the region $a <  \Lambda$ to the region $a \gg \Lambda$, 
where the the transseries would be convergent. 
Then the formal transseries becomes 
a well-defined transseries and 
gives back an analytic function defined 
in Eq.\,\eqref{D2_ON_Exact_Real}:
\begin{eqnarray}
F(s_a)=
2 \int^{s_{a}}_{0} ds \,\frac{\cosh s-1}{s} = 2 {\rm Chin}(s_a).
\label{eq:F(a)_exact}
\end{eqnarray}
After obtaining the analytic function, 
we can safely continue it to the region $a<\Lambda$ 
and find that
\beq
\lim_{a \rightarrow 0} F(s_a) = 0. 
\eeq 
Thus the final result of the $a\to 0$ limit is 
that we can neglect the contribution 
$F(s_a)_{\rm formal}$ altogether, including those 
imaginary ambiguities contained in $F(s_a)_{\rm formal}$. 
We also note that the imaginary ambiguities in the IR contribution 
$F(s_a)_{\rm formal}$ changes from the $\Lambda^8$ term to the $\Lambda^0$ 
term in the process of the analytic continuation to the region $a>\Lambda$. 
It is interesting to note that the function $F(s_a)$ is an 
example of functions of the renormalized coupling $\lambda_a$ 
that can be continued analytically beyond the Landau singularity 
at $a=\Lambda$ to the negative values $\lambda_a<0$ exhibiting 
an entirely different behavior \cite{Yamazaki:2019arj} 
compared to the region $\lambda_a>0$: 
power expandable in $a/\Lambda$ in the region $a<\Lambda$, and 
Borel resummations of divergent power series in $\lambda_a$ as 
coefficients of power series in $\Lambda/a$ in the region $a \gg \Lambda$. 
\newpage 

\section{Two Point Function}
\label{sec:correlation_function}
So far we have seen the resurgence structure of 
the condensate $\langle \delta D^2 \rangle$. 
A similar but more complicated structure can be seen 
in the transseries expansion of the two point function 
\beq
\langle \delta D(x) \delta D(0) \rangle = \int \frac{d^2p}{(2\pi)^2} \, e^{i \mathbf p \cdot \mathbf x} \, \Delta (p).  
\eeq
In the following we assume that $1/x$ is larger than $\Lambda$ ($\Lambda x < 1$) for simplicity. 
A convenient way to obtain 
the transseries expansion of two point function 
is to use its relation to the condensate 
with a UV cutoff $\tilde a$
\beq
\langle \delta D(x) \delta D(0) \rangle = \int_0^\infty d\tilde a \, x J_1(\tilde a x) \langle \delta D^2 \rangle_{\tilde a},
\label{eq:J1_relation}
\eeq
which can be shown by using the property of the Bessel functions $J_l(p x)$ 
\beq
\int \frac{d^2 p}{(2\pi)^2} e^{i \mathbf p \cdot \mathbf x} \, f(p) = \int_0^\infty \frac{dp}{2\pi} J_0(p x) f(p)= \int_0^\infty d\tilde a \left[ x J_1(\tilde a x) \int_0^{\tilde a} \frac{dp}{2\pi} f(p) \right],
\eeq
for any function $f(p)$. 
As in the previous case, 
it is necessary to introduce an IR cutoff $a$ 
to obtain each term in the transseries. 
As we have seen above, 
the transseries for the condensate 
with UV cutoff $\tilde a$ and IR cutoff $a$ can be written as
\beq
\langle \delta D^2 \rangle_{\tilde a,a} = \frac{1}{2} \sum_{l=0}^\infty \Lambda^{2l} \int_0^\infty dt \, \Lambda^t \Big[ \tilde a^{2\eta_l(t)} - a^{2\eta_l(t)} \Big] \frac{P_{l}(t)}{\eta_l(t)}, ~~~\mbox{for} ~~~ \Lambda < a < \tilde a, 
\eeq
where $P_l(t)$ and $\eta_l(t)$ are given by
\beq
P_l(t) = \frac{(-1)^l}{l!} \Big[ (t+l+1)^{(l)} - 4 l (t+l)^{(l-1)} \Big], \hs{10}
\eta_l(t) = 2 - l - \frac{t}{2},
\label{eq:def_P_l}
\eeq
where $(a)^{(l)} = \Gamma(a+l)/\Gamma(a) = a (a+1) \cdots (a+l-1)$ denotes the Pochhammer symbol. 
Note that $P_l(t)$ are polynomials of $t$ and have no singularity. 
From this expression and the relation \eqref{eq:J1_relation}, 
we obtain the transseries for the two point function
with IR cutoff $a > \Lambda$ as
\beq
\langle \delta D(x) \delta D(0) \rangle_a &=& \int_a^\infty d\tilde a \, x J_1(\tilde a x) \langle \delta D^2 \rangle_{\tilde a,a} \label{eq:2pt_func_Borel} \\
&=& \frac{\Lambda^4}{2} \sum_{l=0}^\infty \int_0^\infty dt \, \left( \frac{\Lambda^2 x^2}{4} \right)^{-\eta_l(t)} \left[ \frac{\Gamma(\eta_l(t))}{\Gamma(1-\eta_l(t))} - \mathcal F_l (a x, t) \right] P_{l}(t), \notag 
\eeq
where 
\beq
 \mathcal F_l(a x,t) &=& \frac{1}{\eta_l(t)} \, {}_1 F{}_2 \left( \eta_l(t); 1,1+\eta_l(t),-\frac{a^2x^2}{4} \right) \left( \frac{a^2 x^2}{4} \right)^{\eta_l(t)} \notag \\
&=& \sum_{n=0}^\infty \frac{(-1)^n}{(n!)^2} \frac{1}{\eta_l(t)+n} \left( \frac{a^2 x^2}{4} \right)^{\eta_l(t)+n}.
\eeq
We can show that 
each integrand in Eq.\,\eqref{eq:2pt_func_Borel} 
has no pole on the positive real axis on the complex $t$-plane 
and hence the Borel resummation gives 
a finite result with no ambiguity. 
Therefore, it would be possible to obtain 
a closed form for the two point function with $a=0$ 
by an analytic continuation. 

Next, let us consider what becomes of each term 
in the transseries 
when the IR cutoff $a$ becomes smaller than $\Lambda$. 
To obtain the correct series in each non-perturbative sector 
for $a < \Lambda$, 
the Borel resummation for the $a$-dependent term must 
be performed along the negative real axis of the $t$-plane. 
In other words, the integral must be modified as
\beq
\langle \delta D(x) \delta D(0) \rangle_a = \frac{\Lambda^4}{2} \sum_{l=0}^\infty \int_{-\infty}^\infty dt \left( \frac{\Lambda^2 x^2}{4} \right)^{-\eta_l(t)} \left[ \frac{\Gamma(\eta_l(t))}{\Gamma(1-\eta_l(t))} \theta(t) + \mathcal F_l (a x, t) \theta(-t) \right] P_{l}(t),\notag 
\eeq
where $\theta(t)$ is the step function. 
Since each integrand has singularities 
at points such that $\eta_l(t)=-n~(n=0,1,2,\cdots)$, 
i.e. $t=2(2-l+n)$, 
we need to regularize the integral. 
Although we can regularize the integral
by shifting the integration contour as
${\rm Im} \, t = \pm \epsilon$, 
the result depends on the sign of $t$.
Each singularity gives rise to the ambiguity
\beq
{\rm Im} \, \langle \delta D(x) \delta D(0) \rangle_a = \pm \pi \Lambda^4\sum_{l=0}^\infty \sum_{n=0}^\infty A_{l,n}\left( \frac{\Lambda^2 x^2}{4} \right)^n,
\label{eq:2p_ambiguity}
\eeq
where the coefficients $A_{l,n}$ are given by
\beq
A_{l,n} = (-1)^{l+n} \frac{1}{(n!)^2} \left[ \binom{2n+4}{l} - 4 \binom{2n+2}{l-1} \right], 
\label{eq:coefficient_B_{l,n}}
\eeq
where $\tbinom{p}{q}=\frac{\Gamma(p+1)}{\Gamma(q+1)\Gamma(p-q)}$ denotes the binomial coefficient. 
\begin{table}[t!]
\beq
\begin{array}{c|ccccccccc}
n \backslash l & 0 & 1 & 2 & 3 & 4 & 5 & 6 & 7 & \cdots \\ \hline
0 & -1 & 0 & 2 & 0 & -1 & 0 & 0 & 0 & \cdots \\
1 & 1 & -2 & -1 & 4 & -1 & -2 & 1 & 0 & \cdots \\
~\vdots~ & \phantom{-11} & \phantom{-11} & \phantom{-11} & \phantom{-11} & \phantom{-11} & \phantom{-11} & \phantom{-11} & \phantom{-11} & \phantom{-11}
\end{array} \notag
\eeq
\vs{-5}
\caption{Coefficients $A_{l,n}$.}
\label{table:coefficients}
\end{table}
Summing over $n=0,1,2\cdots$, 
we find that each term in the transseries 
has an ambiguity that is a non-trivial function of $\Lambda x$
\beq
{\rm Im} \, \langle \delta D(x) \delta D(0) \rangle_a \big|_{l=0} = \pm \pi \Lambda^4 J_0(\Lambda x), \hs{5}
{\rm Im} \, \langle \delta D(x) \delta D(0) \rangle_a \big|_{l=1} = \pm \pi \Lambda^5 x J_1(\Lambda x), \hs{5} \cdots, \notag
\eeq
where $J_l(\Lambda x)$ are the Bessel functions. 
For higher $l$, the ambiguities can be determined as follows.
Let $G_l$ and $H_l$ be functions of $\Lambda x$ defined as
\beq
\hs{-5} 
G_l = \Lambda^4 \sum_{n=0}^\infty \frac{(-1)^{l+n} }{(n!)^2} \binom{2n+4}{l} \left( \frac{\Lambda^2 x^2}{4} \right)^n, \hs{1} 
H_l = \Lambda^4 \sum_{n=0}^\infty \frac{(-1)^{l+n} }{(n!)^2} \binom{2n+2}{l-1} \left( \frac{\Lambda^2 x^2}{4} \right)^n. 
\label{eq:def_GH}
\eeq
We can show that these function satisfy the recursion relations
\beq
G_{l+1} = - \frac{1}{l+1} \Big[ \Lambda \p_\Lambda - l \Big] G_l, \hs{10}
H_{l+1} = - \frac{1}{l} \Big[ \Lambda \p_\Lambda - (l+1) \Big] H_l.
\label{eq:recursion_GH}
\eeq
Starting with the initial terms 
$H_0=0$ and $G_0 = -H_1 = \Lambda^4 J_0(\Lambda x)$, 
we can determine $G_l$, $H_l$ and 
the ambiguity of the two point funciton
\beq
{\rm Im} \, \langle \delta D(x) \delta D(0) \rangle_a = \pm \pi \sum_{l=0}^\infty \left( G_l - 4 H_l \right) = \pm \pi \sum_{l=0}^\infty \Lambda^{2l} P_l(\Lambda \p_\Lambda) \Big[ \Lambda^{4-2l} J_0(\Lambda x) \Big] .
\eeq
On the other hand, 
summing over $l=0,1,2,\cdots$ 
and using the binomial theorem 
$\sum_q \tbinom{p}{q} z^q = (1+z)^p$, 
we can show that the total ambiguity cancel 
(see Table \ref{table:coefficients})
\beq
{\rm Im} \, \langle \delta D(x) \delta D(0) \rangle_a = 0. 
\eeq
As in the case of the condensate, 
the singularities on the negative real axis of 
the Borel plane $(t<0)$ are relevant 
to the cancellation of the imaginary ambiguities 
for $a < \Lambda$. 

\section{$\C P^{N-1}$ sigma model}
\label{sec:OtherModels}

It is straightforward to apply our computations above to the ${\mathbb C}P^{N-1}$ sigma model 
\beq
\mathcal L = \frac{1}{g^2} \bigg[ \sum_{a=1}^N |\D_i \phi^a|^2 + D \left( |\phi^a|^2 - 1 \right) \bigg] ,
\eeq
where $D$ is a Lagrange multiplier, 
$\D_i \phi^a = (\p_i + i A_i) \phi^a$ is the covariant derivative
and $A_i$ is an auxiliary $U(1)$ gauge field. 
Here we compute the cancellation of the imaginary ambiguities following Sec.\,\ref{sec:Cancellation}.  

Integrating out the complex scalar fields $\phi^a$ 
with the ansatz $A_\mu = 0$ and $D={\rm const.}$, 
we obtain the same effective potential as \eqref{V_eff2}, 
whose minimum is given by 
\beq
\langle D \rangle = \Lambda^2 = \mu^2 e^{-\frac{4\pi}{\lambda_\mu}},  
\eeq
where $\lambda_\mu = g_\mu^2 N$ is the 't Hooft coupling renormalized at $\mu$. Like the $O(N)$ sigma model, the theory is asymptotically free, and the mass gap at large $N$ is identical to that in the $O(N)$ model in Eq.\,(\ref{Lambda}).  

In addition to the condensate of the auxiliary field 
$\langle \delta D^2 \rangle$, 
we can consider the condensate of field strength,
which takes the form 
\begin{equation}
\left< F_{\mu \nu} ^2 \right>
=
- \frac{8\pi}{N}
\int \frac{d^2 p}{(2\pi)^2} \, p^2 \sqrt{\frac{p^2}{ p^2 + 4 \Lambda^2}} \, \frac{1}{s_p} + \mathcal O(N^{-2}),
\label{Gluon_Condensate}
\end{equation}
with $s_p = 4 \, {\rm arcsinh} (p/2\Lambda)$. 
We can explicitly perform the integral to obtain 
the exact expression in the large-$N$ limit
\beq
\left< F_{\mu \nu}^2 \right>_{\tilde a} = \frac{2}{N} \Lambda^4 \left[ 4 {\rm Chin} \Big( \frac{s_{\tilde a}}{2} \Big) - {\rm Chin} (s_{\tilde a}) \right],
\label{eq:exact_cp}
\eeq
where $\tilde a$ is the UV cutoff, and ${\rm Chin}(x)$ is the entire function defined by the integral in \eqref{D2_ON_Exact_Real}. 

On the other hand, the transseries expression with an IR cufoff $a > \Lambda$ takes the form
\beq
\left< F_{\mu \nu}^2 \right>_{\tilde a,a} &=& - \frac{1}{2N} \sum_{l=0}^\infty \Lambda^{2l} \int_0^\infty dt \, \Lambda^t \Big[ \tilde a^{2\eta_l(t)} - a^{2\eta_l(t)} \Big] \frac{\tilde P_{l}(t)}{\eta_l(t)}, 
\eeq
where $\tilde P_l(t)$ and $\eta_l(t)$ are given by
\beq
\tilde P_{l}(t) = \frac{(-1)^l }{\Gamma(l+1)}\frac{\Gamma (t+2l+1)}{\Gamma(t+l+1)}, \hs{10}
\eta_l(t) = 2-l-\frac{t}{2}.
\eeq
For $a > \Lambda$, there is no singularity 
on the positive real axis on the Borel plane and 
hence the exact expression \eqref{eq:exact_cp} can be obtained 
by an analytic continuation to $a \rightarrow 0$. 

If we consider the continuation of the transseries to the region where $a < \Lambda$, the Borel resummation should be modified as 
\beq
\left< F_{\mu \nu}^2 \right>_{\tilde a,a} &=& - \frac{1}{2N} \sum_{l=0}^\infty \Lambda^{2l} \int_{-\infty}^\infty dt \, \Lambda^t \Big[ \tilde a^{2\eta_l(t)} \theta(t) - a^{2\eta_l(t)} \theta(-t) \Big] \frac{P_{l}(t)}{\eta_l(t)}
\label{eq:ts_continued_cp}
\eeq
In this case, the terms with $l=0,1,3,4$ have 
imaginary ambiguities associated with the poles at 
\beq
t = 4 - 2l ~~~ (l=0,1,3,4).   
\eeq 
The term with $l=2$ also has an ambiguity since it contains $\log \lambda_{\tilde a}$ and $\log \lambda_a$.
Although each term has an imaginary ambiguities, 
$\left< F_{\mu \nu}^2 \right>$ has no imaginary part due to the cancellation 
\beq
 {\rm Im} \left< F_{\mu \nu}^2 \right>_{\tilde a,a}
 =
 \frac{\pm \pi}{N}
\bigg[ 
\left( \tilde a e^{-\frac{2\pi}{\lambda_{\tilde a}}} \right)^4
-
4 \left( \tilde a e^{-\frac{2\pi}{\lambda_{\tilde a}}} \right)^2 \Lambda^2
+
6 \Lambda^4
-
4 \left( \tilde{a} e^{-\frac{2\pi}{\lambda_{\tilde{a}}}} \right)^{-2} \Lambda^6
+
\left( \tilde{a} e^{-\frac{2\pi}{\lambda_{\tilde{a}}}} \right)^{-4} \Lambda^8
\bigg] = 0. \notag
\eeq

We next look at the compactified model on ${\mathbb R} \times S^1$ with the $\Z_N$ symmetric twisted boundary conditions. We first take the circumference of the compactified dimension $L$ small $L \Lambda \ll 1$ but fixed in the large $N$ limit $NL\Lambda \gg 1$.  This conventional large-$N$ limit is different from the Abelian large $N$ limit $NL\Lambda \ll 1$ where the monopole-instantons can be computed \cite{Poppitz:2012nz}.
We impose the twisted boundary conditions on the field as
\begin{equation}
\phi^a (x_1+n L, x_2) = e^{in L m^a} \phi^a (x_1, x_2), ~~~\mbox{with $n \in \mathbb{R}$ and $m^a = 2 \pi a / (NL)$},
\end{equation}
where the coordinates of ${\mathbb R}$ and $S^1$ are denoted by $x_1$ and $x_2$, respectively. 
We set the periodic boundary conditions 
for the auxiliary field $D$ and the gauge field. 
The effective action for the auxiliary field $D$ is given as
\begin{eqnarray}
V_{\rm eff} (D)
=
\frac{1}{2 L} \sum^N_{a=1} \sum_{n \in \mathbb{Z}} \int_{\mathbb{R}} \frac{dp_2}{2 \pi} \log \Big[ \left( k^a_n \right)^2 + p^2_2 + D \Big]
- \frac{D}{2 g^2},
\end{eqnarray}
where the Matsubara frequency is given as $k^a_n = 2\pi n/L + m^a$.
At large $N$, we obtain the same effective action as ${\mathbb R}^2$:
\begin{eqnarray}
V_{\rm eff} (D)
&=&
 \frac{N}{2} \frac{1}{ N L}  \sum_{n \in \mathbb{Z}} \int_{\mathbb{R}} \frac{dp_2}{2 \pi} \log \left[ \left( \frac{2 \pi n}{NL}\right)^2 + p^2_2 + D \right]
- \frac{D}{2 g^2}
\nonumber \\
&\rightarrow&
\frac{N}{2} \int_{\mathbb{R}^2} \frac{d^2p}{\left( 2 \pi \right)^2} \log \left( p^2 + D \right)
- \frac{D}{2 g^2} . \phantom{\Bigg|}
 \label{Veff_LargeN}
\end{eqnarray}
This is a consequence of the volume independence at large $N$ \cite{Sulejmanpasic:2016llc}.
Therefore the gap equation is unchanged, and we obtain the same mass gap as before in Eq.\,(\ref{Lambda}).  

To compute the condensate, 
we need to write the momentum integral in  (\ref{Gluon_Condensate}) as
\begin{equation}
\left< F_{\mu \nu}^2 \right>
=
- \frac{8\pi}{NL}  \sum_{n \in \mathbb{Z}}
\int^{\tilde{a}}_{a} \frac{dp_2}{2\pi} p^2 \sqrt{\frac{p^2}{p^2 + 4 \Lambda^2}} \, \frac{1}{s_p} + \mathcal O(N^{-2}),\end{equation}
with $p = \sqrt{(2\pi n/L)^2+p_2^2}$.
We still need the momentum cutoff $a$ due to the IR divergence in the semiclassical expansion. 
To compute the imaginary ambiguities for small $L$, 
we only need to look at the zero Matsubara mode, 
because the nonzero Matsubara mode acts as a large momentum cutoff for and eliminates the pole in the  momentum integral.   
Following Sec.\,\ref{sec:Cancellation}, we can compute the imaginary ambiguities as
\begin{eqnarray}
&&
{\rm Im} \left< \delta D^{2} \right>^{{\mathbb R} \times S^1}_{\tilde a,a}
\\
&&
=
\frac{\pm \pi }{L}  \left[ 
2\left( \tilde a e^{- \frac{2 \pi}{\lambda_{\tilde a}}}  \right)^3
+
2 \left( \tilde a e^{- \frac{2 \pi}{\lambda_{\tilde a}}}  \right) \Lambda^2
-
2  \left( \tilde a e^{- \frac{2 \pi}{\lambda_{\tilde a}}}  \right)^{-1} \Lambda^4
-
2 \left( \tilde a e^{- \frac{2 \pi}{\lambda_{\tilde a}}}  \right)^{-3} \Lambda^6
\right] \theta(\Lambda-a),\nonumber
\end{eqnarray}
for the condensate 
in the $O(N)$ sigma model while
\begin{eqnarray}
&&
{\rm Im} \left< F_{\mu \nu}^2 \right>^{{\mathbb R}\times S^1}_{\tilde a,a} \label{ImF_RXS}
\\
&&
=
  \frac{\pm \pi}{NL}  \left[ 
2\left( \tilde a e^{- \frac{2 \pi}{\lambda_{\tilde a}}}  \right)^3
-
6 \left( \tilde a e^{- \frac{2 \pi}{\lambda_{\tilde a}}}  \right) \Lambda^2
+
6  \left( \tilde a e^{- \frac{2 \pi}{\lambda_{\tilde a}}}  \right)^{-1} \Lambda^4
-
2 \left( \tilde a e^{- \frac{2 \pi}{\lambda_{\tilde a}}}  \right)^{-3} \Lambda^6
\right]
\theta(\Lambda-a),\nonumber
\end{eqnarray}
for the condensate in ${\mathbb C}P^{N-1}$ model. 
They both vanish but have a different structure 
than the case of ${\mathbb R}^2$.
The first term in Eq.\,(\ref{ImF_RXS}) is computed in Ref.\,\cite{Ishikawa:2019tnw} and they agree. 

In the calculation above, 
we have assumed that the only zero mode 
is relevant to the imaginary ambiguity. 
However, we have to take into account 
the contributions of the higher Matsubara modes 
to see how the results in the compact 
and non-compact cases are related. 
For that purpose, it is convenient to consider 
the imaginary ambiguity of the correlation functions. 
For the two point function of the fluctuations of the auxiliary field, 
it is convenient to use the Poisson resummation formula
\beq
\sum_{n \in \Z} f(2\pi n /L) = \sum_{\nu \in \Z} \frac{1}{L} \int \frac{dp}{2\pi} e^{i p \nu L} f(p). 
\eeq
The two point function of the auxiliary field $\delta D$ 
in the compactified case is given by
\begin{equation}
\left< \delta D(x) \delta D(0) \right>_a = 
8\pi \sum_{\nu \in \Z}
\int^{\infty}_{a} \frac{d^2p}{(2\pi)^2} \, e^{i p (x+\nu L)} \frac{\sqrt{p^2 \left( p^2 + 4 \Lambda^2 \right)}}{s_p} + \mathcal O(N^{-1}),
\end{equation}
where we have fixed the position of 
the first operator $\delta D(x)$ 
at $(x_1,x_2)=(x,0)$ for simplicity.
From the ambiguity of the two point function on $\R^2$ in \eqref{eq:2p_ambiguity}, 
we obtain the ambiguity of the
 $\mathcal O(\Lambda^l)$ term for $a < \Lambda$ as 
\beq
{\rm Im} \langle \delta D(x) \delta D(0) \rangle_a \big|_{l} &=& \pm \pi \sum_{\nu \in \Z} \Big[ G_l(x+\nu L) - 4 H_l(x+\nu L) \Big], 
\eeq
where the functions $G_l$ and $H_l$ are defined in \eqref{eq:def_GH}. 
By using the Poisson resummation formula, 
the summation over the integer $\nu$ 
can be rewritten back into the Kaluza Klein momentum number $n$. 
For example, $G_0=-H_1=\Lambda^4 J_0(\Lambda x)$ can be rewritten as
\beq
\pi \sum_{\nu \in \Z} G_0(x+\nu L) = - \pi \sum_{\nu \in \Z} H_1(x+\nu L) = \frac{\Lambda^3}{R} \sum_{n \in \Z} \frac{e^{-i \frac{n}{R}x}}{\sqrt{1-\frac{n^2}{R^2 \Lambda^2}}} \, \theta \! \left(\Lambda^2 - \frac{n^2}{R^2} \right),
\eeq 
where $R=L/2\pi$ is the compactification radius.
The higher order terms can also be determined by using 
the recursion relation \eqref{eq:recursion_GH} as 
\beq
{\rm Im} \, \langle \delta D(x) \delta D(0) \rangle_a \big|_{l} &=& \pm \pi \sum_{n \in \Z} \Lambda^{2l} P_l(\Lambda \p_\Lambda) \left[  \frac{\Lambda^{3-2l}}{R} \frac{e^{-i \frac{n}{R}x}}{\sqrt{1-\frac{n^2}{R^2 \Lambda^2}}} \, \theta \! \left(\Lambda^2 - \frac{n^2}{R^2} \right)\right],
\label{eq:ambiguity_2pt_compact}
\eeq 
where $P_l(t)$ is the polynomial given in Eq.\,\eqref{eq:def_P_l}. 
The step function $\theta(\Lambda^2- n^2/R^2)$
in the the imaginary ambiguity \eqref{eq:ambiguity_2pt_compact} 
implies that Stokes phenomena occur 
every time one of Kaluza Klein masses (Matsubara frequencies) 
$n/R$ becomes smaller than the scale $\Lambda$. 
In particular, the ambiguity of the perturbative part ($l=0$) 
changes from $\mathcal O(\Lambda^3/R)$ to $\mathcal O(\Lambda^4)$
due to the infinitely many Stokes phenomena which occur 
as the compactification radius $R$ is varied from zero to infinity
\beq
{\rm Im} \, \langle \delta D(x) \delta D(0) \rangle_a \big|_{l=0} = \pm
 \begin{cases} 
\Lambda^3/R & \mbox{for $R < \Lambda^{-1}$} \\ \Lambda^4 + \cdots & \mbox{for $R \rightarrow \infty$}
\end{cases}.
\eeq 
This explains the discrepancy between 
the imaginary ambiguities in the models 
on ${\mathbb R}^2$ and $\R \times S^1$ 
with the $\Z_n$ twisted boundary condition \cite{Ishikawa:2019tnw}. 

\section{Conclusions}
\label{sec:Conclusions}

We have studied the resurgence structure of 
the condensate and two-point function
in the $O(N)$ sigma model at large $N$ 
using the semi-classical expansion.  
We have deduced the semi-classical ansatz in Eq.\,(\ref{eq:condensate_CF_p}) 
from the exact solution at large $N$ by using an expansion in 
powers of $\Lambda^2/p^2$ and 
a small-coupling $\lambda_{p}$ expansion 
before performing the momentum integral.  
The expansion suffers from the renormalon and IR divergences, both 
of which are typical in the semiclassical expansion in QFT.  

In order to circumvent the IR problem, 
we have introduced the IR cutoff $a$ in the momentum integral.  The imaginary ambiguities arise 
when the cutoff is small, $a< \Lambda$.  
We have identified the known renormalon ambiguity 
of the condensate at the leading order $\Lambda^0$
in the semiclassical expansion (\ref{Delta_deltaD}), 
and show that it is cancelled only 
after computing up to order $\Lambda^8$, 
rather than $\Lambda^4$ as previously thought.  
We have also examined the result 
in Eq.\,\eqref{CF_SC_Real} of the semiclassical ansatz 
comparing it to the exact solution. 
We find that the result of the semiclassical ansatz 
can be understood in terms of the transseries expansion 
of the exact result at large $N$. 
To understand the behavior as a function of the IR cutoff $a$, 
we have first taken the transseries expansion 
of the condensate in powers of $\Lambda^{2}/a^{2}$ and $\lambda_a$. 
Their coefficients $c_{\left(2l, n\right)}$ at order $\Lambda^{2l}/a^{2l-4}$ and 
$\lambda^n_a$ can be analytically continued in 
$\lambda_a$ to the region $a<\Lambda$ where $\lambda_a<0$, 
which reproduces the result of the semiclassical ansatz. 
In this way, the imaginary ambiguity at order $\Lambda^{8}/a^4$ 
can be understood as coming from the Borel resummation of 
the IR contribution. 
We have also found that the limit of $a\to 0$ of the result of 
the semiclassical ansatz can be taken 
if we first make an analytic continuation to $a \gg \Lambda$ and 
then sum over $l$ (the Borel resummation) and over $n$ of the 
transseries to obtain an analytic function, which leads to 
the correct $a\to 0$ limit. 
With this procedure we have been able to recover the exact result at large $N$.


We have computed the cancellation of the renormalon in the semiclassical expansion in other models, such as the $O(N)$ and ${\mathbb C}P^{N-1}$ models on ${\mathbb R} \times S^1$ at large $N$. They turn out to be more complicated as multiple nonperturbative sectors give rise to imaginary ambiguities. 
In particular, we found that 
there exist infinitely many Stokes phenomena 
which occur as the compactification radius $R$ is
varied from zero to infinity. 
It is important to further investigate these models in the future.


\begin{acknowledgements}
This work is supported by the Ministry of Education, Culture, 
Sports, Science, and Technology (MEXT)-Supported Program for the 
Strategic Research Foundation at Private Universities ``Topological 
Science" (Grant No. S1511006) 
and 
by the Japan Society for the Promotion of Science (JSPS) 
Grant-in-Aid for Scientific Research (KAKENHI) Grant Number 
(18H01217).
This work is also supported in part by JSPS KAKENHI Grant Numbers 
18K03627, 21K03558 (T.\ F.) and 19K03817 (T.\ M.).
\end{acknowledgements}


\appendix

\section{Generic structure of Borel transform for condensate}
\label{appendix:general_Borel}
In this appendix, we derive the Borel transforms
for the condensate \eqref{eq:general_condensate}
\beq
\langle \mathcal O(0)^2 \rangle |_{a,\tilde a} 
=
\int_{|a|<|p|<|\tilde a|} \frac{d^dp}{(2\pi)^d} \, \Delta(p) \notag = 
\sum_{l=0}^\infty \int_{|a|<|p|<|\tilde a|} \frac{d^dp}{(2\pi)^d} \, p^{[\Delta]} \left( \frac{\Lambda}{p} \right)^{\beta_0 \sigma_l} f_l (\lambda_p'),
\label{eq:condensate_general}
\eeq
where $\lambda_p' = 2\pi/(\beta_0 \log \frac{p}{\Lambda})$. 
By the change of variable, 
\beq
p = \Lambda \left(\frac{\mu}{\Lambda} \right)^{\upsilon}, \hs{5}
\left( \upsilon = \frac{\lambda_\mu'}{\lambda_p'} = \frac{\log p/\Lambda}{\log \mu / \Lambda} \right),
\eeq
the condensate \eqref{eq:condensate_general} can be rewritten as
\beq
 \langle \mathcal O(0)^2 \rangle |_{\tilde a,a} = C \, \Lambda^{2[\mathcal O]} \, \sum_{l=0}^\infty \int_{\upsilon_a}^{\upsilon_{\tilde a}} d\upsilon \, \left( \frac{\Lambda}{\mu} \right)^{e_l \, \upsilon} f_l(\lambda_\mu'/\upsilon)
~~~~ \mbox{with} ~~~~ C = \frac{d \log \frac{\mu}{\Lambda}}{(4\pi)^\frac{d}{2} \Gamma(d/2+1)},
\label{eq:Borel_expansion}
\eeq
where 
\beq
e_l = \beta_0 \sigma_l-2[\mathcal O], \hs{5} 
\upsilon_p = \frac{\log p/\Lambda}{\log \mu/\Lambda}.
\eeq 
To rewrite \eqref{eq:Borel_expansion} 
into a Borel resummed form, 
let us decompose the integral as
\beq
\int_{\upsilon_a}^{\upsilon_{\tilde a}} d\upsilon \, \left( \frac{\Lambda}{\mu} \right)^{e_l \, \upsilon} f_l(\lambda_\mu'/\upsilon) &=& \int_{\upsilon_a}^{\infty} d\upsilon \, \left( \frac{\Lambda}{\mu} \right)^{e_l \, \upsilon} f_l(\lambda_\mu'/\upsilon) - \int_{\upsilon_{\tilde a}}^{\infty} d\upsilon \, \left( \frac{\Lambda}{\mu} \right)^{e_l \, \upsilon} f_l(\lambda_\mu'/\upsilon).
\eeq
Then, by change of variables $\upsilon=t/e_l +\upsilon_a$ and $\upsilon=t/e_l + \upsilon_{\tilde a}$, we obtain 
\beq
\Lambda^{2[\mathcal O]} \int_{\upsilon_a}^{\upsilon_{\tilde a}} d\upsilon \, \left( \frac{\Lambda}{\mu} \right)^{e_l \, \upsilon} f_l(\lambda_\mu'/\upsilon) &=& \mu^{2[\mathcal O]} \left( \frac{\Lambda}{\mu} \right)^{\beta_0 \sigma_l} \int_{0}^{\infty} dt \, \left( \frac{\Lambda}{\mu} \right)^{t} B_l(t),
\eeq
with
\beq
B_l(t) = \frac{1}{e_l} \left[ \left( \frac{\mu}{a} \right)^{e_l} f_l \left( \frac{e_l \lambda'_\mu}{t + e_l \upsilon_a} \right) - \left( \frac{\mu}{\tilde a} \right)^{e_l} f_l \left(\frac{e_l \lambda'_\mu}{t+e_l \upsilon_{\tilde a}} \right) \right]. 
\eeq

\section{Perturbative expansions}
\label{sec:perturbation}

In this appendix, we explicitly calculate the expansion coefficients $c_{(2l,n)}$.  
Defining $c_{\left(2l, n \right)}= c_{\left(2l, n \right)}(\tilde a) - c_{\left(2l, n \right)}(a)$, we have
\begin{eqnarray}
c_{(0,n)} (p)
&=&
\int dp \frac{2 p^3 t^n_p}{\left( 4 \pi \right)^{n+1} } \nonumber
\\
&=&
\frac{\tilde{a}^4}{\left(  8 \pi \right)^{n+1}} \Gamma(n+1, 2 t_p)
\label{c_0n}
\\
c_{(2,n)} (p)
&=&
\int dp \frac{4p^2 \left(  t^n_p - n t^{n-1}_p\right)}{\left( 4\pi \right)^{n+1}}
 \nonumber\\
&=&
\frac{2 p^2 t^n_p}{\left(4 \pi \right)^{n+1}}
\\
c_{(4,n)} (p)
&=&
\int dp \frac{2 \left(  -2 t^n_p - n t^{n-1}_p +2 \left( n \right)_2 t^{n-2}_p\right)}{\left( 4\pi \right)^{n+1} p}
 \nonumber\\
&=&
\frac{ 2t^{n+1}_p }{\left( 4\pi \right)^{n+1} \left( n+1 \right)}
+
\frac{ t^n_p - 2 n t^{n-1}_p}{\left( 4\pi \right)^{n+1} }
\\
c_{(6,n)} (p)
&=&
\int dp \frac{
4 \left(
6 t^{n}_p + 5 n t^{n-1}_p   - 3 \left(n\right)_2 t^{n-2}_p -2 \left(n\right)_3 t^{n-3}_p  
\right)}{3 \left( 4 \pi \right)^{n+1}  p^3}
 \nonumber\\
&=&
\frac{2 \left( -6 t^n_p 
+ n t^{n-1}_p
+ 2 \left( n \right)_2 t^{n-2}_p
\right)}{3 \left( 4 \pi \right)^{n+1} p^2 }
\\
c_{(8,n)} (p)
&=&
\int dp \frac{
-60 t^{n}_p -59 n t^{n-1}_p +11 \left( n \right)_{2} t^{n-2}_p +20 \left( n\right)_{3} t^{n-3}_p + 4 \left( n\right)_{4} t^{n-4}_p
}{3 \left( 4 \pi \right)^{n+1}  p^5}
 \nonumber\\
&=&
\frac{\left( -1 \right)^{n+1}}{ \tilde{a}^4 \left( 8 \pi \right)^{n+1}}  \Gamma(n+1, -2 t_p)
+ \frac{33 t^{n}_p + 13 n t^{n-1} -12 \left( n \right)_{2} t^{n-2}_p -4 \left( n \right)_{3} t^{n-3}_p }{6 \left( 4 \pi \right)^{n+1} p^4},
\label{c8_p}
\end{eqnarray}
where $t_p = \log \left(\tilde{a}^2 / p^2 \right)$ and $\left( n \right)_k = n\left(n-1\right) \left(n-2\right) \dots \left(n-k+1\right)$ is the falling factorial. 
To compute the integrals, we have used the property of the incomplete Gamma function
\beq
\Gamma(n+1,z) = n \Gamma(n,z) + z^n e^{-z}.
\eeq

To sum $C_{2l} = \sum^{\infty}_{n=0} \lambda^{n+1} c_{(2l,n)}$, we set $x = \lambda t_p/ 4 \pi$ and use
\begin{equation}
\sum^{\infty}_{n=0} \frac{x^{n+1}}{n+1}
=
-\log(1-x)
\;\;\;\;\;
\mbox{and}
\;\;\;\;\;
\frac{d^k}{dx^k } x^n = \left( n \right)_{k} x^{n-k} .
\end{equation}   
We find that they are equivalent to the ones computed in Sec.\,\ref{sec:PerturbativeExpansion}, as they should, up to a constant for $C_4$.

\section{Power series expansion in $\Lambda^2$}
\label{sec:power_series}

In this appendix, we show the transseries expansion of the condensate. 
We first expand the variable $u_{\tilde a}$ in powers of $\Lambda^2/\tilde a^2$ as 
\begin{equation}
u_{\tilde a}
=-4\sum_{k=1}^\infty\frac{1}{k}\left(\sum_{l=1}^\infty\frac{(2l-3)!!}{l! 2}
\left(\frac{-2\Lambda^2}{\tilde a^2}\right)^l\right)^k. 
\label{eq:u_mu_power_series}
\end{equation}
Summation over $l$ is convergent when $2\Lambda/\tilde a<1$, and the 
sum over $k$ is also convergent when $\sqrt{1+(4\Lambda^2/\tilde a^2)}<2$. 

We can expand the function $F(s_{\tilde a})$ around $F(8\pi/\lambda(\tilde a))$ 
in powers of $u_{\tilde a}=s_{\tilde a}-8\pi/\lambda(\tilde a)$ as 
\begin{eqnarray}
&&F(s_{\tilde a})-F\left(\frac{8\pi}{\lambda_{\tilde a}}\right)
=\int_{\frac{8\pi}{\lambda_{\tilde a}}}^{\frac{8\pi}{\lambda_{\tilde a}}+u_{\tilde a}}ds
\frac{e^s-2+e^{-s}}{s}
\label{eq:power_series_F_smu}
\\
&=&
\sum_{m=1}^{\infty}\frac{(u_{\tilde a})^{m}}{m}\left[
\frac{\tilde a^4}{\Lambda^4}
\sum_{n=0}^{m-1}
\frac{(-1)^n}{(m-n-1)!}
\left(\frac{\lambda_{\tilde a}}{8\pi}\right)^{n}
+2\left(-\frac{\lambda_{\tilde a}}{8\pi}\right)^{m}
-
\frac{\Lambda^4}{\tilde a^4}
\sum_{n=0}^{m-1}
\frac{(-1)^m}{(m-n-1)!}
\left(\frac{\lambda_{\tilde a}}{8\pi}\right)^{n}
\right]. 
\nonumber
\end{eqnarray}
The radius of convergence is given by $|u_{\tilde a}|<8\pi/\lambda_{\tilde a}$. 
Therefore we can obtain the transseries expansion of the 
condensate using the power series expansion 
\eqref{eq:power_series_F_smu} together with the Borel resummed 
transseries expansion of $F(8\pi/\lambda_{\tilde a})$ in 
\eqref{eq:Borel_resum_mu}.

To find the power series expansion in $a/\Lambda$ of the function 
$F(s_a)$ for the IR contribution of the condensate, we first expand 
$F(s_a)$ in powers of $s_a$ and then $s_a$ in powers of $a/\Lambda$ as 
\begin{eqnarray}
F\left(s_a\right)
&=&\int^{s_{a}}_{0} ds \left[ \frac{e^{s}-1}{s}+\frac{e^{-s}-1}{s}\right]
=\sum_{k=1}^\infty\frac{1}{(2k)! k}s_a^{2k}
\label{eq:integral_rep_Fa}
\\
&=&\sum_{k=1}^\infty\frac{1}{(2k)! k}\left\{
\sum_{m=1}^\infty\frac{4(-1)^{m-1}}{m}
\left(\frac{a}{2\Lambda}+\sum_{l=1}^\infty \frac{(-1)^{l-1}(2l-3)!!}{l! 2^l}
\left(\frac{a}{2\Lambda}\right)^{2l}\right)^m\right\}^k
,
\nonumber
\end{eqnarray}
where the sum over $m$ is convergent if $a<2\Lambda$, the sum over 
$l$ is convergent if $\sqrt{1+a^2/(2\Lambda)^2}+a/(2\Lambda)< 2$.



\end{document}